\documentclass[twocolumn]{aastex631}

\newcommand{\um}{$\mu$m}

\newcommand{\htwo}{\ensuremath{\mathrm{H_2}}}
\newcommand{\hone}{\ion{H}{1}}
\newcommand{\hplus}{\ion{H}{2}}

\newcommand{\qpah}{\ensuremath{q_\mathrm{PAH}}}

\newcommand{\ipah}{\ensuremath{I^\mathrm{PAH}_{\rm F770W}}}
\newcommand{\ssfr}{\ensuremath{\mathrm{SFR}/M_\star}}
\newcommand{\mstar}{\ensuremath{M_\star}}
\newcommand{\norm}{\ensuremath{C_\mathrm{F770W}^\mathrm{PAH}}}

\newcommand{\mjysr}{MJy~sr$^{-1}$}

\newcommand{\red}[1]{#1}

\begin{document}

\title{Polycyclic Aromatic Hydrocarbon and CO~(2-1) Emission at $50{-}150$~pc Scales in 70 Nearby Galaxies}
\shorttitle{I can't believe it's not a gas map!}
\shortauthors{Chown et al.}

\correspondingauthor{Ryan Chown}
\email{rchown53@gmail.com}

\newcommand{\Ox}{Sub-department of Astrophysics, Department of Physics, University of Oxford, Keble Road, Oxford OX1 3RH, UK}

\newcommand{\UGent}{Sterrenkundig Observatorium, Universiteit Gent, Krijgslaan 281 S9, B-9000 Gent, Belgium}

\newcommand{\STScI}{Space Telescope Science Institute, 3700 San Martin Drive, Baltimore, MD 21218, USA}

\newcommand{\IWR}{Universit\"{a}t Heidelberg, Interdisziplin\"{a}res Zentrum f\"{u}r Wissenschaftliches Rechnen, Im Neuenheimer Feld 205, 69120 Heidelberg, Germany}
\newcommand{\Radcliffe}{Elizabeth S. and Richard M. Cashin Fellow at the Radcliffe Institute for Advanced Studies at Harvard University, 10 Garden Street, Cambridge, MA 02138, U.S.A.}

\newcommand{\MPIA}{Max-Planck-Institut f\"{u}r Astronomie, K\"{o}nigstuhl 17, D-69117, Heidelberg, Germany}

\newcommand{\AURA}{AURA for the European Space Agency (ESA), Space Telescope Science Institute, 3700 San Martin Drive, Baltimore, MD 21218, USA}

\newcommand{\UCSD}{Department of Astronomy \& Astrophysics, University of California, San Diego, 9500 Gilman Dr., La Jolla, CA 92093, USA}

\newcommand{\Whitman}{Whitman College, 345 Boyer Avenue, Walla Walla, WA 99362, USA}

\newcommand{\JHU}{Department of Physics and Astronomy, The Johns Hopkins University, Baltimore, MD 21218, USA}

\newcommand{\OSU}{Department of Astronomy, The Ohio State University, 140 West 18th Avenue, Columbus, OH 43210, USA}

\newcommand{\OSUPhys}{Department of Physics, The Ohio State University, Columbus, Ohio 43210, USA}

\newcommand{\CCAPP}{Center for Cosmology and Astroparticle Physics (CCAPP), 191 West Woodruff Avenue, Columbus, OH 43210, USA}

\newcommand{\ARI}{Astronomisches Rechen-Institut, Zentrum f\"{u}r Astronomie der Universit\"{a}t Heidelberg, M\"{o}nchhofstr. 12-14, D-69120 Heidelberg, Germany}

\newcommand{\ANU}{Research School of Astronomy and Astrophysics, Australian National University, Canberra, ACT 2611, Australia}

\newcommand{\UConn}{Department of Physics, University of Connecticut, 196A Auditorium Road, Storrs, CT 06269, USA}

\newcommand{\UHawaii}{Institute for Astronomy, University of Hawaii, 2680 Woodlawn Drive, Honolulu, HI 96822, USA}

\newcommand{\UniCA}{Universit\'{e} C\^{o}te d'Azur, Observatoire de la C\^{o}te d'Azur, CNRS, Laboratoire Lagrange, 06000, Nice, France}

\newcommand{\UAlberta}{Dept. of Physics, University of Alberta, 4-183 CCIS, Edmonton, Alberta, T6G 2E1, Canada}

\newcommand{\Arcetri}{INAF — Osservatorio Astrofisico di Arcetri, Largo E. Fermi 5, I-50125, Florence, Italy}

\newcommand{\UWyoming}{Department of Physics and Astronomy, University of Wyoming, Laramie, WY 82071, USA}

\newcommand{\LJMU}{Astrophysics Research Institute, Liverpool John Moores University, 146 Brownlow Hill, Liverpool L3 5RF, UK}

\newcommand{\ITA}{Universit\"{a}t Heidelberg, Zentrum f\"{u}r Astronomie, Institut f\"{u}r Theoretische Astrophysik, Albert-Ueberle-Str 2, D-69120 Heidelberg, Germany}

\newcommand{\CfA}{Center for Astrophysics $\mid$ Harvard \& Smithsonian, 60 Garden St., 02138 Cambridge, MA, USA}

\newcommand{\MPE}{Max-Planck-Institut f\"{u}r Extraterrestrische Physik (MPE), Giessenbachstr. 1, D-85748 Garching, Germany}

\newcommand{\UMD}{Department of Astronomy and Joint Space-Science Institute, University of Maryland, College Park, MD 20742, USA}

\newcommand{\UVA}{Department of Astronomy, University of Virginia, Charlottesville, VA, USA}

\newcommand{\NRAO}{National Radio Astronomy Observatory, Charlottesville, VA, USA}

\newcommand{\ASIAA}{Institute of Astronomy and Astrophysics, Academia Sinica, No. 1, Sec. 4, Roosevelt Road, Taipei 106216, Taiwan}

\newcommand{\kipac}{Kavli Institute for Particle Astrophysics \& Cosmology (KIPAC), Stanford University, CA 94305, USA}

\newcommand{\aifa}{Argelander-Institut f\"ur Astronomie, University of Bonn, Auf dem H\"ugel 71, 53121 Bonn, Germany}

\newcommand{\TKU}{Department of Physics, Tamkang University, No.151, Yingzhuan Road, Tamsui District, New Taipei City 251301, Taiwan}

\newcommand{\CarnegieObs}{The Observatories of the Carnegie Institution for Science. 813 Santa Barbara Street, Pasadena, CA 91101, USA}

\newcommand{\Princeton}{Department of Astrophysical Sciences, Princeton University, Princeton, NJ 08544, USA}

\newcommand{\IAS}{Institute for Advanced Study, 1 Einstein Drive, Princeton, NJ 08540, USA}

\newcommand{\COOL}{Cosmic Origins Of Life (COOL) Research DAO, coolresearch.io}

\newcommand{\ESO}{European Southern Observatory (ESO), Karl-Schwarzschild-Stra{\ss}e 2, 85748 Garching, Germany}

\newcommand{\ULyon}{Univ Lyon, Univ Lyon 1, ENS de Lyon, CNRS, Centre de Recherche Astrophysique de Lyon UMR5574, F-69230 Saint-Genis-Laval, France}

\newcommand{\UoM}{Jodrell Bank Centre for Astrophysics, Department of Physics and Astronomy, University of Manchester, Oxford Road, Manchester M13 9PL, UK}

\author[0000-0001-8241-7704]{Ryan~Chown}
\affiliation{\OSU}

\author[0000-0002-2545-1700]{Adam~K.~Leroy}
\affiliation{\OSU}
\affiliation{\CCAPP}

\author[0000-0002-4378-8534]{Karin~Sandstrom}
\affiliation{\UCSD}

\author[0000-0002-5235-5589]{J\'{e}r\'{e}my~Chastenet}
\affiliation{\UGent}

\author[0000-0002-9183-8102]{Jessica~Sutter}
\affiliation{\Whitman}
\affiliation{\UCSD}

\author[0000-0001-9605-780X]{Eric W. Koch}
\affiliation{\CfA}

\author[0009-0001-5949-1524]{Hannah B. Koziol}
\affiliation{\UCSD}

\author[0000-0001-9793-6400]{Lukas~Neumann}
\affiliation{\aifa}
\affiliation{\ESO}

\author[0000-0003-0378-4667]{Jiayi Sun}
\altaffiliation{NASA Hubble Fellow}
\affiliation{\Princeton}

\author[0000-0002-0012-2142]{Thomas G. Williams}
\affiliation{\Ox}

\author[0000-0003-4974-3481]{Dalya Baron}
\affiliation{\CarnegieObs}
\affiliation{\kipac}

\author[0000-0002-5259-2314]{Gagandeep S. Anand}
\affiliation{\STScI}

\author[0000-0003-0410-4504]{Ashley.~T.~Barnes}
\affiliation{\ESO}

\author[0009-0001-1221-0975]{Zein~Bazzi}
\affiliation{\aifa}

\author[0000-0002-2545-5752]{Francesco Belfiore}
\affiliation{\Arcetri}

\author[0000-0003-0166-9745]{Frank Bigiel}
\affiliation{\aifa}

\author[0000-0002-5480-5686]{Alberto Bolatto}
\affiliation{\UMD}

\author[0000-0003-0946-6176]{Médéric~Boquien}
\affiliation{\UniCA}

\author[0000-0001-5301-1326]{Yixian Cao}
\affiliation{\MPE}

\author[0000-0002-5635-5180]{M\'elanie Chevance}
\affiliation{\ITA, \COOL}

\author[0000-0001-6498-2945]{Dario Colombo}
\affiliation{\aifa}

\author[0000-0002-5782-9093]{Daniel~A.~Dale}
\affiliation{\UWyoming}

\author[0000-0002-8760-6157]{Jakob~den~Brok}
\affiliation{\CfA}

\author[0000-0002-4755-118X]{Oleg V. Egorov}\affiliation{\ARI}

\author[0000-0002-1185-2810]{Cosima Eibensteiner}
\altaffiliation{Jansky Fellow of the National Radio Astronomy Observatory}
\affiliation{\NRAO}

\author[0000-0002-6155-7166]{Eric Emsellem}
\affiliation{\ESO}
\affiliation{\ULyon}

\author[0000-0002-8806-6308]{Hamid Hassani}
\affiliation{\UAlberta}

\author[0000-0001-9656-7682]{Jonathan D. Henshaw}
\affiliation{\LJMU}
\affiliation{\MPIA}

\author[0000-0001-9020-1858]{Hao He}
\affiliation{\aifa}

\author[0000-0002-0432-6847]{Jaeyeon Kim}
\affiliation{\kipac}

\author[0000-0002-0560-3172]{Ralf~S.~Klessen}
\affiliation{\ITA}
\affiliation{\IWR}
\affiliation{\CfA}
\affiliation{\Radcliffe}

\author[0000-0001-6551-3091]{Kathryn Kreckel}
\affiliation{\ARI}

\author[0000-0003-3917-6460]{Kirsten~L.~Larson}
\affiliation{\AURA}

\author[0000-0003-0946-6176]{Janice~C.~Lee}
\affiliation{\STScI}

\author[0000-0002-6118-4048]{Sharon E. Meidt}
\affiliation{\UGent}

\author[0000-0001-7089-7325]{Eric J. Murphy}
\affiliation{\NRAO}

\author[0000-0002-0119-1115]{Elias K. Oakes}
\affiliation{\UConn}

\author[0000-0002-0509-9113]{Eve C. Ostriker}
\affiliation{\Princeton}
\affiliation{\IAS}

\author[0000-0002-1370-6964]{Hsi-An Pan}
\affiliation{\TKU}

\author[0000-0003-2721-487X]{Debosmita~Pathak}
\affiliation{\OSU}
\affiliation{\CCAPP}

\author[0000-0002-5204-2259]{Erik Rosolowsky}
\affiliation{\UAlberta}

\author[0000-0002-2545-1700]{Sumit K. Sarbadhicary}
\affiliation{\OSU}
\affiliation{\CCAPP}
\affiliation{\JHU}

\author[0000-0002-3933-7677]{Eva Schinnerer}
\affiliation{\MPIA}

\author[0000-0003-4209-1599]{Yu-Hsuan Teng}
\affiliation{\UMD}

\author[0000-0002-8528-7340]{David A. Thilker}
\affiliation{\JHU}

\author[0009-0005-8923-558X]{Tony D. Weinbeck}
\affiliation{\UWyoming}

\author[0000-0002-7365-5791]{Elizabeth J. Watkins}
\affiliation{\UoM}

\begin{abstract}
Combining Atacama Large Millimeter/sub-millimeter Array CO(2-1) mapping and JWST near- and mid-infrared imaging, we characterize the relationship between CO(2-1) and polycyclic aromatic hydrocarbon (PAH) emission at $\approx 100$~pc resolution in 70 nearby star-forming galaxies. Leveraging a new Cycle 2 JWST treasury program targeting nearby galaxies, we expand the sample size by more than an order of magnitude compared to previous $\approx 100$~pc resolution CO-PAH comparisons. Focusing on regions of galaxies where most of the gas is likely to be molecular, we find strong correlations between CO(2-1) and 3.3~$\mu$m, $7.7\mu$m, and $11.3~\mu$m PAH emission, estimated from JWST's F335M, F770W, and F1130W filters. We derive power law relations between CO(2-1) and PAH emission, which have indices in the range $0.8{-}1.3$, implying relatively weak variations in the observed CO-to-PAH ratios across the regions that we study. We find that CO-to-PAH ratios and scaling relationships near \ion{H}{2} regions are similar to those in  diffuse sight lines. The main difference between the two types of regions is that sight lines near \ion{H}{2} regions show higher intensities in all tracers. Galaxy centers, on the other hand, show higher overall intensities and enhanced CO-to-PAH ratios compared to galaxy disks. Individual galaxies show $0.19$~dex scatter in the normalization of CO at fixed $I_{\rm PAH}$, and this normalization anti-correlates with specific star formation rate (\ssfr) and correlates with stellar mass. We provide a prescription that accounts for these galaxy-to-galaxy variations and represents our best current empirical predictor to estimate CO(2-1) intensity from PAH emission, which allows one to take advantage of JWST's excellent sensitivity and resolution to trace cold gas.
\end{abstract}

\keywords{Interstellar medium (847), Dust continuum emission (412), CO line emission (462), Disk galaxies (391), Dust nebulae (413), Extragalactic astronomy (506)}

\section{Introduction} \label{sec:intro}

Broad emission features at 3.3, 6.2, 7.7, 8.6, 11.2, and 12.7~\um, attributed to the stretching and bending modes of polycyclic aromatic hydrocarbons \citep[PAHs;][]{puget1989, leger1984, allamandola1989, tielens2008}, dominate the near- and mid-infrared (NIR and MIR) luminosities of star-forming galaxies \citep[e.g.,][]{smith2007,tielens2008,galliano2018}. Emission from PAHs has been used to trace the star formation rate \citep{peeters2004, calzetti2007,belfiore2023}, interstellar radiation field, and PAH abundance \citep[e.g.,][]{draine2007,kennicutt2012,whitcomb2023,baron2024,sutter2024}. At large scales (from a few kpc to integrated galaxies), PAH emission also shows a close correspondence with CO emission, exhibiting strong correlations and nearly linear scaling relations between PAH and CO intensity over three orders of magnitude \citep[see][]{regan2004,gao2019,chown2021,gao2022,leroy2021,leroy2023a,whitcomb2023}. This correlation resembles the one that has been observed for decades relating tracers of molecular gas and the overall dust content traced by extinction, mm-wave emission, and far-infrared emission 
\citep[e.g.,][]{hildebrand1983,israel1997,leroy2011,bolatto2013, galliano2018}. 
CO emission traces the molecular gas in galaxies \citep[e.g.,][]{bolatto2013}, and this has led to the suggestion that PAH emission may be used as a quantitative tracer of the interstellar medium (ISM), specifically molecular gas in star-forming galaxies. This prospect is particularly exciting because the widespread availability of maps tracing PAH emission from \textit{WISE}, \textit{Spitzer}, and now JWST, means that high-resolution, high-sensitivity maps of the atomic and molecular ISM could potentially be produced from every image of PAH emission obtained using these telescopes. In principle this could be done to higher redshifts than can be accessed with CO observations. Furthermore, the upcoming Spectro-Photometer for the History of the Universe, Epoch of Reionization and Ices Explorer satellite \citep[SPHEREx,][]{dore2018}, which will provide full-sky spectral maps of PAH emission, could also in principle be used to make maps of the  atomic and molecular ISM across all nearby galaxies.

Initial JWST imaging of nearby galaxies supports the idea that PAH emission may act as a cold, dense ISM tracer. JWST images using PAH-dominated filters resemble sharper, more sensitive versions of Atacama Large Millimeter/sub-millimeter Array (ALMA) CO maps for the same galaxies \citep[][]{leroy2023,sandstrom2023a}. These analyses showed an excellent quantitative correspondence between CO~(2-1) and emission in the F770W and F1130W filters\footnote{The F770W filter is centered at $7.7\mu$m and is $1.95\mu$m wide. The F1130W filter is centered at $11.3\mu$m and is $0.73\mu$m wide.} at $\approx 100$~pc scales \citep{leroy2023a}, even suggesting that the PAH emission traces lower-density ISM emission into the \ion{H}{1}-dominated parts of the ISM \citep{sandstrom2023a}. However, these first studies focused on only four galaxies, and so the general relationship between CO and PAH emission at these 50{-}150~pc scales (similar to the sizes of molecular clouds, or ``cloud scales'') was not statistically robust. Fortunately, over its first two cycles of operations, JWST has nearly completed an extensive census of NIR and MIR PAH emission from $74$ local ($D \lesssim 20$~Mpc) galaxies as part of the Physics at High Angular Resolution in Nearby Galaxies Survey (PHANGS) Cycle 1 and Cycle 2 treasuries (see Appendix \ref{app:treasury}). In addition to  $0.03\arcsec$ to $0.8\arcsec$ ($\approx 15$ to $120$~pc) resolution JWST images, all of these galaxies also have $\approx 1\arcsec$ ($\approx 50$ to $150$~pc) resolution CO~(2-1) imaging from ALMA \citep{leroy2021}.

In this paper, the first to take full advantage of the combined PHANGS-JWST Cycle 1 and 2 surveys, we use this large new set of ALMA and JWST data to make quantitative comparisons between CO(2-1) and PAH emission traced using JWST's F770W, F335M\footnote{The F335M filter is centered at 3.362~\um\ and is 0.347\um\ wide.}, and F1130W filters. We consider the 70 nearby galaxies with current in-hand JWST and high-quality ALMA observations, and address the following questions:

\begin{enumerate}
\item What are the median and scatter in the CO(2-1)/PAH band ratios across a representative sample of nearby star-forming galaxies at molecular cloud scales (50 to 150~pc)? What are the correlation strengths and best-fit parameters for the CO(2-1) vs. PAH power law scaling relationships?

\item Are there significant differences between the correlations measured for galaxy centers and galaxy disks, or when contrasting emission near \ion{H}{2} regions with diffuse emission? Such variations might be expected given the different distributions of MIR intensity from these regions \citep{pathak2024} and the observation of systematic suppression of PAH emission from inside \ion{H}{2} regions \citep{pety2005, lebouteiller2007, compiegne2008, chastenet2023,egorov2023,sutter2024}.

\item How does the observed CO(2-1)-PAH relationship vary from galaxy to galaxy? Does it change as a function of parameters that also correlate with dust and PAH abundances or the interstellar radiation field? Or do these factors affect CO emission in a similar way to PAH emission, suppressing the impacts of environmental variations on the correlation?
\end{enumerate}

We describe our expectations for this work in \S\ref{sec:motivation}, our approach in \S \ref{sec:data}, present our results in \S \ref{sec:results}, and discuss and summarize our conclusions in \S \ref{sec:discussion} and \S \ref{sec:conclusions}. Throughout, we focus on F770W emission because it has the widest availability out of all of the PAH bands that we consider, in terms of the total number of galaxies observed and total area mapped.

\section{Expectations}\label{sec:motivation}

Throughout the analysis, we will reference the expectation for emission from stochastically-heated PAHs subjected to a scaled version of the local interstellar radiation field and mixed with an ISM consisting of mostly molecular gas \citep[e.g.,][]{draine2007a,draine2011,compiegne2010}. To first order we expect
\begin{eqnarray}
\label{eq:dust}
 I_{\rm PAH} &\propto& \mathrm{DGR}\times \qpah  \times N_{\rm H_2}\times U \\
\nonumber &\propto&  \left(\mathrm{DGR}\times \qpah  \times X_\mathrm{CO}\times U \right)I_{\rm CO},
\end{eqnarray}
where $I_{\rm PAH}$ and $I_{\rm CO}$ are the observed intensities of PAH and CO emission respectively, DGR is the dust-to-gas mass ratio, \qpah\ is the PAH-to-dust mass fraction, $U$ is the strength of the interstellar radiation field relative to that in the Solar neighborhood \citep[\qpah\ and $U$ are defined in][]{draine2007}, and $X_{\rm CO}$ is the CO-to-\htwo\ conversion factor. 

This equation indicates why we expect a scaling relation in the first place. It also indicates the factors that might lead to environmental variations in the $I_{\rm CO}/I_{\rm PAH}$ ratio, and non-linearities in the observed scaling relations, e.g. when $U\gtrsim 10^3$ \citep[for more discussion see][]{draine2007a, leroy2023a,leroy2023,sandstrom2023a}. For example, to first order, $U$ is proportional to the local unextinguished $\Sigma_\mathrm{FUV} \propto \Sigma_\mathrm{SFR}$ \citep[e.g., see recent simulations by][]{linzer2024}, and so we expect PAH emission to be brighter where local star formation activity is more intense (though bear in mind the destruction of PAHs in \ion{H}{2} regions mentioned above). We return to this expected correlation between $I_{\rm PAH}/I_{\rm CO}$ and star formation activity in \S\ref{sec:galtogal}. The hardness of the radiation field, variations in the dust-to-gas ratio, and PAH abundance variations may also be important, but in our selected molecular gas dominated regions of relatively massive galaxies, we mostly expect these to be second-order effects (see \S \ref{sec:galtogal}). 

\section{Data and Methods}
\label{sec:data}

We analyze 70 galaxies that have both high-resolution ALMA CO~(2-1) imaging and JWST imaging tracing PAH emission. Our targets are all part of the PHANGS surveys, and as such are relatively massive ($M_\star \gtrsim 10^{9.5}~\mathrm{M_\odot}$), star-forming (SFR/$M_\star \gtrsim 10^{-11}~\mathrm{yr^{-1}}$), and moderate inclination ($i \lesssim 70^\circ$) galaxies within $D \lesssim 20$~Mpc \citep{leroy2021}. The subset of 19 targets presented in \citet{lee2023} and \citet{williams2024} have both F1130W imaging and ``nebular masks,'' which identify \ion{H}{2} regions based on a morphological decomposition of the H$\alpha$ emission observed by the Very Large Telescope's Multi-Unit Spectroscopic Explorer \citep[VLT/MUSE][]{groves2023}. The remaining targets are part of a new Cycle 2 treasury program, which we describe below and in Appendix \ref{app:treasury}. These new targets lack F1130W imaging and are not systematically covered with optical integral field spectroscopy. They therefore lacking nebular masks. Both the JWST Cycle 1 and Cycle 2 data include the F335M filter but the continuum subtraction needed to isolate the PAH emission has only reached a science-ready state for the Cycle 1 data (\S\ref{subsec:jwst}). Therefore analyses that require nebular masks (\S~\ref{sec:nebular}) and/or F335M and/or F1130W are limited to our 19 Cycle 1 targets.

\subsection{JWST imaging of PAH emission}\label{subsec:jwst}

We use JWST MIRI and NIRCam imaging of 19 galaxies from the PHANGS-JWST Cycle 1 Treasury \citep[GO~2107, PI: J. Lee;][]{lee2023}. These data cover the F335M, F770W, and F1130W bands, each of which captures a strong PAH feature \citep[e.g.,][]{tielens2008}. The F335M filter captures the prominent 3.3~\um\ feature, attributed to C--H stretching modes of small PAHs, while the F770W filter captures the 7.7~\um\ PAH feature, attributed to C--C stretching modes of larger PAHs, and the F1130W filter captures the 11.3~\um\ feature, attributed to C--H out-of-plane bending modes of larger PAHs \citep{tielens2008}. 

We also analyze 51 galaxies from a new Cycle 2 Treasury (GO~3707, PI: A. Leroy), which we describe in Appendix \ref{app:treasury}. These new observations include the F335M and F770W filters. From the 54 available Cycle 2 galaxies, we remove NGC~3344 due to the lack of CO~(2-1) observations, NGC~5530 because the observations are still to be executed, and NGC~1068 because the background level in the JWST images remain too uncertain for the current analysis. In total we analyze 70 galaxies. The distribution of gas-phase metallicities of these galaxies has median $12+\log\mathrm{O/H}=8.49$~dex and robust standard deviation of 0.09~dex. %

Reduction for both data sets was done using the PHANGS-JWST pipeline (\texttt{pjpipe}\footnote{\url{https://pjpipe.readthedocs.io/en/latest/}}), following \citet{williams2024}. \texttt{pjpipe} is a wrapper on \texttt{jwst}\footnote{\url{https://jwst-pipeline.readthedocs.io/en/latest/index.html}}, tailored to produce mosaics of images with extended emission. 

The Rayleigh-Jeans tail of the stellar continuum contributes to F335M and F770W (more so for F335M). This contribution must be subtracted from the total surface brightnesses to obtain emission from PAHs. We calculate stellar-continuum-subtracted F770W surface brightness $I_{\rm F770W}^{\rm PAH}$ by subtracting F200W (Cycle 1) or F300M (Cycle 2) times a scaling factor (one factor per band) following \citet{sutter2024}. In our sample, the F770W stellar continuum correction (F770W$_\star = 0.22\times$F300M, or $0.13\times$F200W for Cycle 1) tends to be largest in the gas-poor, high stellar surface density inner regions of galaxies, including bars and bulges, which have high stellar-to-gas ratios. The coverage of F300M and F200W is smaller than that of F770W, and so for pixels without F300M or F200W data we subtract median(F770W$_\star$/F770W)$\times $F770W, where the median \citep[which is 2--10\%, in agreement with ][]{whitcomb2023a} is computed separately for each galaxy. 

For F335M data, the stellar continuum subtraction is critical to any estimate of PAH emission \citep{sandstrom2023}. We use the continuum subtraction method defined by  H. Koziol et al. (in preparation) based on \citet{sandstrom2023}, where the F300M and F360M (for Cycle 1) are used to estimate continuum. We denote the continuum-subtracted F335M surface brightness as $I_{\rm F335M}^{\rm PAH}$. A similar effort using the F300M is underway for Cycle 2 but not yet ready, and so our analysis of $I_{\rm F335M}^{\rm PAH}$ focused only on the Cycle 1 targets.

The F770W filter also captures emission from hot, very small dust grains, but this has a small effect \citep[in the 19 PHANGS-JWST Cycle 1 galaxies, 7.7~\um\ PAH emission is found to be about $5\times$ brighter than the underlying continuum;][]{baron2024a} and we do not subtract any underlying dust continuum. In both data sets, we also masked out a few bright stars and background galaxies with elliptical apertures that cover all of the emission from these sources on top of the F770W images. We do not use any data within these apertures.

\subsection{ALMA CO(2-1)}\label{subsec:alma}

We use PHANGS--ALMA CO~(2-1) observations, which are described in \citet{leroy2021}. All maps include short- and zero-spacing data and are expected to achieve full flux recovery. The native resolution is $\approx 1.0\arcsec$, varying slightly from galaxy to galaxy. We use a noisy but uniform and high-completeness version of the CO~(2-1) maps to ensure that $>90\,\%$ of the CO flux enters our analysis and that our mean trends will be unbiased by signal-to-noise based clipping of the CO data \citep[see][]{leroy2023}. These ``flat'' maps are produced following \citet{neumann2023} using a procedure analogous to spectral stacking \citep{schruba2011}. For each line of sight, we integrate over a fixed-width velocity window around the local mean velocity defined via either low-resolution CO emission, H$\alpha$ emission, \ion{H}{1} emission, or a model of the circular rotation, according to whichever yielded the most complete coverage and coherent reference velocity field for each galaxy. The velocity window is adapted to the disk-average line width of each individual galaxy and can vary from 20 to 200\,km/s. We additionally ensure that all significant CO emission is included by combining the fixed-width velocity mask with the ``strict'' mask from \citet{leroy2021}. These ``flat'' masks will capture all of the CO emission along each line of sight and have well-defined noise, making them ideal to calculate the mean CO~(2-1) intensity, $I_{\rm CO}$, as a function of $I_{\rm PAH}$.

\subsection{Environment masks}\label{subsec:masks} 

We analyze how the CO-to-PAH correlations vary as functions of environment within galaxies, focusing on three specific environments: 1) galaxy centers; 2) regions outside galaxy centers with prominent nebular emission (``nebular regions''), which are $\gtrsim 80\%$ \ion{H}{2} regions but also include a $\sim 10\%$ contribution from supernova remnants \citep[][]{li2024}; and 3) regions outside of centers and not covered by the nebular region mask (``diffuse regions''). Following \citet{pathak2024} these three regions show distinct distributions of mid-IR intensity. These three regions likely show differences in radiation field strength and/or hardness, PAH abundance, \htwo/\hone\ ratio, and dust-to-gas ratio. Additionally, PAH emission is found to be systematically suppressed inside \ion{H}{2} regions \citep{chastenet2023,egorov2023,sutter2024,chown2024}. 

We define galaxy centers using the masks in \citet{querejeta2021}, which are based on near-IR stellar morphology, and nebular regions following \citet{santoro2022, groves2023}, which are based on ionized gas emission observed by the VLT/MUSE as part of the PHANGS-MUSE survey \citep[][]{emsellem2022}. The galaxy center masks are available for all targets, though only 53 of our targets have well-defined central regions. The rest of the targets do not have any pixels classified as being in a ``center'', and all of their pixels are included in the ``disk'' category. The nebular masks are available only for the $19$ Cycle 1 targets, which restricts a subset of our analysis to that subsample. See \citet{pathak2024} for further details on a similar application of these environment masks.

\subsection{Analysis approach}
\label{subsec:approach}

To quantify the observed relationship between CO and PAH emission, we compile all measurements of individual pixels for all 70 galaxies into a single table, with columns for each NIR and MIR band (\S \ref{subsec:jwst}), CO(2-1) intensity (\S \ref{subsec:alma}), and the values of each mask (\S \ref{subsec:masks}). To avoid introducing systematic uncertainties, we analyze CO(2-1) intensities and do not adopt a CO-to-H$_2$ conversion factor. In Appendix~\ref{appendix:howto} we describe how to use our results to estimate H$_2$ surface densities.

The CO~(2-1) data have the coarsest angular resolution in this work. The CO resolution varies slightly from galaxy to galaxy. Over the range of distances of the galaxies in the sample, the working resolution for PHANGS--ALMA corresponds to median $98$~pc with a $16{-}84\%$ range of $63{-}129$~pc \citep[Table 15 in][]{leroy2021}. Using \texttt{webbpsf}\footnote{\url{https://webbpsf.readthedocs.io/en/latest/}}-generated JWST PSFs, we convolve all JWST PAH images to share the same resolution of the corresponding CO~(2-1) data following \citet{aniano2011} and \citet{williams2024}. Then we define an astrometric grid with pixels $1/3\times$ the CO PSF FWHM in size so that approximately 9 pixels represent an independent measurement. We reproject the CO~(2-1) intensity (in K~km~s$^{-1}$), JWST PAH intensity (in MJy~sr$^{-1}$), and mask information onto this grid, noting whether the majority of the area in each pixel corresponds to a galaxy center, a nebular region, or neither. We correct all intensities to face-on values by scaling by $\cos i$, where $i$ is the galaxy inclination from \citet[][]{lang2020} and/or \citet{leroy2021}. The sample consists of nearly-face-on galaxies (median $\cos i = 0.7$), and so this correction does not have a large impact on our analysis. 

PAH emission is also expected to emerge from dust mixed with atomic gas \citep[e.g., see][]{sandstrom2023a}. To avoid lines of sight where \hone\ is expected to make up most of the ISM, we consider only regions that have inclination-corrected $I_\mathrm{F770W}^{\rm PAH} \geq 0.5~\mathrm{MJy~sr^{-1}}$ \citep[see][for arguments about this specific value for ``bright'' emission]{leroy2023a}. In Table~\ref{tab:fracs} we report the percentages of flux or area that are classified as ``bright''  according to this criterion. Focusing on F770W-bright pixels means that we analyze the significant majority of flux in all bands across the sample. However, we do exclude a majority ($49\%$) of the area in the maps outside galaxy centers and nebulae. Analysis of this faint, extended diffuse emission requires the inclusion of \ion{H}{1} and will be the topic of future work.

The $3.3~\mu$m PAH feature is $>10\times$ fainter than the 7.7 and 11.3~$\mu$m ones \citep[e.g.,][]{chastenet2023,dale2025}, and the maps tend to be more limited by noise and systematic uncertainties related to continuum subtraction. As a result, we impose an additional threshold for analyzing the $3.3~\mu$m emission, considering only lines of sight with $I_{\rm F335M}^{\rm PAH} > 0.1$~MJy~sr$^{-1}$. This corresponds roughly to $I_{\rm F770W}^{\rm PAH} \gtrsim 2$~MJy~sr$^{-1}$ and was selected to catch the lines of sight where the $3.3~\mu$m map yields robust detections at resolution matched to ALMA. Our analysis of F335M images is limited to lines of sight with surface brightness above this threshold.

\begin{deluxetable}{lrrrrr}\label{tab:fracs}
 \tablecaption{Fraction of flux and area entering analysis}
    \tablehead{
         \colhead{Quantity} & \colhead{All} & \colhead{Outside centers} & \colhead{Centers} & \colhead{Nebular} & \colhead{Diffuse} \\
        \colhead{} & \colhead{\%} & \colhead{\%} & \colhead{\%} & \colhead{\%} & \colhead{\%}}
\startdata
F335M$_\mathrm{PAH}$\tablenotemark{a} & 100 & 100 & 100 & 100 & 100 \\
F770W$_\mathrm{PAH}$\tablenotemark{a} & 94 & 93 & 100 & 99 & 89 \\
F770W & 93 & 91 & 100 & 99 & 86 \\
F1130W & 93 & 92 & 100 & 99 & 88 \\
CO(2-1) & 96 & 95 & 100 & 99 & 94 \\
\hline
Area & 51 & 51 & 90 & 82 & 47 \\
\enddata
\tablenotetext{a}{F335M$_\mathrm{PAH}$ and F770W$_\mathrm{PAH}$ refer to intensity in those filters after stellar continuum subtraction. After this correction, both filters are expected to be dominated by PAH emission (\S \ref{subsec:approach}).}
\tablecomments{Percentages of the flux or area in the full map captured in the subset of F770W-bright pixels with $I_\mathrm{F770W_{PAH}}\geq 0.5$~MJy~sr$^{-1}$ (see \S \ref{sec:data}), by band and environment. Our analysis captures the majority of the flux in all bands, but only includes about half the total observed area.}
\end{deluxetable}

Using the matched-resolution measurements, we analyze the correlation between CO~(2-1) emission and emission in PAH-dominated JWST filters. For each MIR band $X$, we record the Spearman rank correlation coefficient ($r$), as well as the median and scatter in the ratio $I_{\rm CO(2-1)}/I_{\rm X}$. We then calculate a best-fit power law relating $I_{\rm CO(2-1)}$ to $I_{\rm X}$. Because the PHANGS--ALMA CO maps are much less sensitive than the JWST F770W and F1130W maps at matched angular resolution \citep[see][]{leroy2023}, we treat the mid-IR imaging as the independent ($x$) axis for this calculation. We construct logarithmically-spaced bins in $I_\nu^\mathrm{X}$ and then compute the median and scatter, captured by the 16{-}84\% range, of $I_\mathrm{CO(2-1)}$ within each bin.

We perform linear regression on these binned measurements\footnote{Because of the large number of individual pixels, the mean CO~(2-1) intensity is detected at high signal-to-noise in all bins.} using \texttt{linmix}\footnote{\url{https://linmix.readthedocs.io/en/latest/index.html}} a hierarchical Bayesian method described in \cite{kelly2007}. It performs a linear regression of $y$ on $x$ while incorporating measurement errors in both variables. We model the CO-vs-PAH emission relationship as
\begin{equation}
\label{eq:line}
\log_{10} I_\mathrm{CO(2-1)}=m(\log_{10} I_\nu^\mathrm{X}-x_0)+b,
\end{equation}
where the pivot $x_0\equiv \mathrm{median}(x)$. Re-centering the fit at $x_0$ ensures minimal covariance between the best-fit $m$ and $b$.

\section{Results}\label{sec:results}

\begin{deluxetable*}{lrrrrrrrr}
\label{tab:fits}
\tablecolumns{8}
\tablecaption{Ratios, correlation, and scaling relations between PAH and CO~(2-1) emission. Each section of the table reports results for a different data selection.}

\tablehead{
\colhead{$X$} & \colhead{$N_\mathrm{gal}$} & \colhead{$N_\mathrm{pix}$\tablenotemark{a}} & \colhead{$\log_{10}\mathrm{CO}/X$} & \colhead{$r$} & \colhead{$b$} & \colhead{$m$} & \colhead{$x_0$ } & \colhead{$\sigma$}}
\startdata
\cutinhead{All pixels}
F335M$_\mathrm{PAH}$ & 19 & 296834 & $1.27 \pm 0.38$ & $0.58$ & $1.38 \pm 0.07$ & $1.01 \pm 0.10$ & $0.10$ & $0.44$ \\
F770W$_\mathrm{PAH}$ & 70 & 2090731 & $0.03 \pm 0.35$ & $0.64$ & $1.39 \pm 0.06$ & $0.90 \pm 0.07$ & $1.44$ & $0.43$ \\
F770W & 70 & 2120025 & $-0.00 \pm 0.35$ & $0.64$ & $1.41 \pm 0.07$ & $0.91 \pm 0.07$ & $1.47$ & $0.43$ \\
F1130W & 20 & 972892 & $-0.12 \pm 0.37$ & $0.63$ & $1.20 \pm 0.08$ & $1.02 \pm 0.09$ & $1.34$ & $0.44$ \\
\cutinhead{All pixels outside of centers}
F335M$_\mathrm{PAH}$ & 19 & 279910 & $1.25 \pm 0.38$ & $0.56$ & $1.17 \pm 0.07$ & $1.04 \pm 0.12$ & $-0.05$ & $0.44$ \\
F770W$_\mathrm{PAH}$ & 70 & 2041245 & $0.02 \pm 0.35$ & $0.62$ & $1.12 \pm 0.07$ & $0.98 \pm 0.08$ & $1.16$ & $0.42$ \\
F770W & 70 & 2069574 & $-0.01 \pm 0.35$ & $0.62$ & $1.14 \pm 0.07$ & $0.99 \pm 0.08$ & $1.19$ & $0.42$ \\
F1130W & 20 & 946437 & $-0.12 \pm 0.37$ & $0.62$ & $0.93 \pm 0.08$ & $1.05 \pm 0.13$ & $1.11$ & $0.44$ \\
\cutinhead{All pixels in centers}
F335M$_\mathrm{PAH}$ & 17 & 16915 & $1.63 \pm 0.34$ & $0.80$ & $1.65 \pm 0.07$ & $0.78 \pm 0.09$ & $0.10$ & $0.41$ \\
F770W$_\mathrm{PAH}$ & 58 & 49469 & $0.18 \pm 0.33$ & $0.88$ & $1.52 \pm 0.07$ & $0.80 \pm 0.07$ & $1.44$ & $0.42$ \\
F770W & 58 & 50434 & $0.11 \pm 0.34$ & $0.87$ & $1.51 \pm 0.07$ & $0.83 \pm 0.07$ & $1.47$ & $0.44$ \\
F1130W & 18 & 26443 & $-0.02 \pm 0.30$ & $0.90$ & $1.30 \pm 0.07$ & $0.97 \pm 0.08$ & $1.34$ & $0.39$ \\
\cutinhead{All pixels in nebular regions (Cycle 1 only)}
F335M$_\mathrm{PAH}$ & 19 & 119051 & $1.19 \pm 0.33$ & $0.70$ & $1.13 \pm 0.07$ & $1.09 \pm 0.12$ & $-0.05$ & $0.39$ \\
F770W$_\mathrm{PAH}$ & 19 & 186605 & $-0.02 \pm 0.34$ & $0.75$ & $1.02 \pm 0.09$ & $1.03 \pm 0.11$ & $1.09$ & $0.40$ \\
F770W & 19 & 187137 & $-0.04 \pm 0.34$ & $0.75$ & $1.03 \pm 0.08$ & $1.06 \pm 0.11$ & $1.12$ & $0.40$ \\
F1130W & 19 & 194681 & $-0.17 \pm 0.33$ & $0.76$ & $0.89 \pm 0.08$ & $1.06 \pm 0.12$ & $1.11$ & $0.40$ \\
\cutinhead{All pixels in diffuse regions (Cycle 1 only)}
F335M$_\mathrm{PAH}$ & 19 & 160845 & $1.31 \pm 0.40$ & $0.44$ & $0.95 \pm 0.12$ & $1.28 \pm 0.32$ & $-0.40$ & $0.46$ \\
F770W$_\mathrm{PAH}$ & 19 & 632119 & $0.08 \pm 0.38$ & $0.52$ & $0.72 \pm 0.13$ & $1.20 \pm 0.27$ & $0.66$ & $0.47$ \\
F770W & 19 & 642906 & $0.05 \pm 0.38$ & $0.52$ & $0.74 \pm 0.13$ & $1.21 \pm 0.28$ & $0.70$ & $0.47$ \\
F1130W & 19 & 727342 & $-0.11 \pm 0.38$ & $0.54$ & $0.68 \pm 0.12$ & $1.22 \pm 0.25$ & $0.81$ & $0.47$ \\
\enddata
\tablenotetext{a}{Pixels have size FWHM/\red{3}, so that there are \red{nine} pixels per independent measurement.}
\tablenotetext{b}{This scatter reflects the combined noise in the CO data, galaxy-to-galaxy scatter, and scatter about the fit within each galaxy. It is usually dominated by the noise in the CO data.}
\tablecomments{
Columns: $X$ --- JWST band compared to CO(2-1) \red{where the subscript ``PAH" indicates that stellar continuum has been subtracted using overlapping NIRCam observations (\S\ref{subsec:jwst})}; $N_{\rm gal}$ --- number of galaxies entering this analysis; $N_{\rm pix}$ --- number of sightlines entering the correlation analysis; $\log_{10} \mathrm{CO}/X$ --- $\log_{10}$ of median ratio of CO(2-1) in K km s$^{-1}$ to intensity in MJy sr$^{-1}$ in band $X$ with error indicating the scatter in the ratio estimated from the median absolute deviation; $r$ --- rank correlation between CO(2-1) and intensity in band $X$ for all sightlines; $b$, $m$, $x_0$ --- best fit power law scaling parameters following Equation~\ref{eq:line} relating CO(2-1) to intensity in band $X$; $\sigma$ --- rms scatter in dex of individual data about the best fit scaling relation, inferred via the median absolute deviation. 
}

\end{deluxetable*}

\begin{figure*}
\begin{center}
\includegraphics[width=0.5\textwidth]{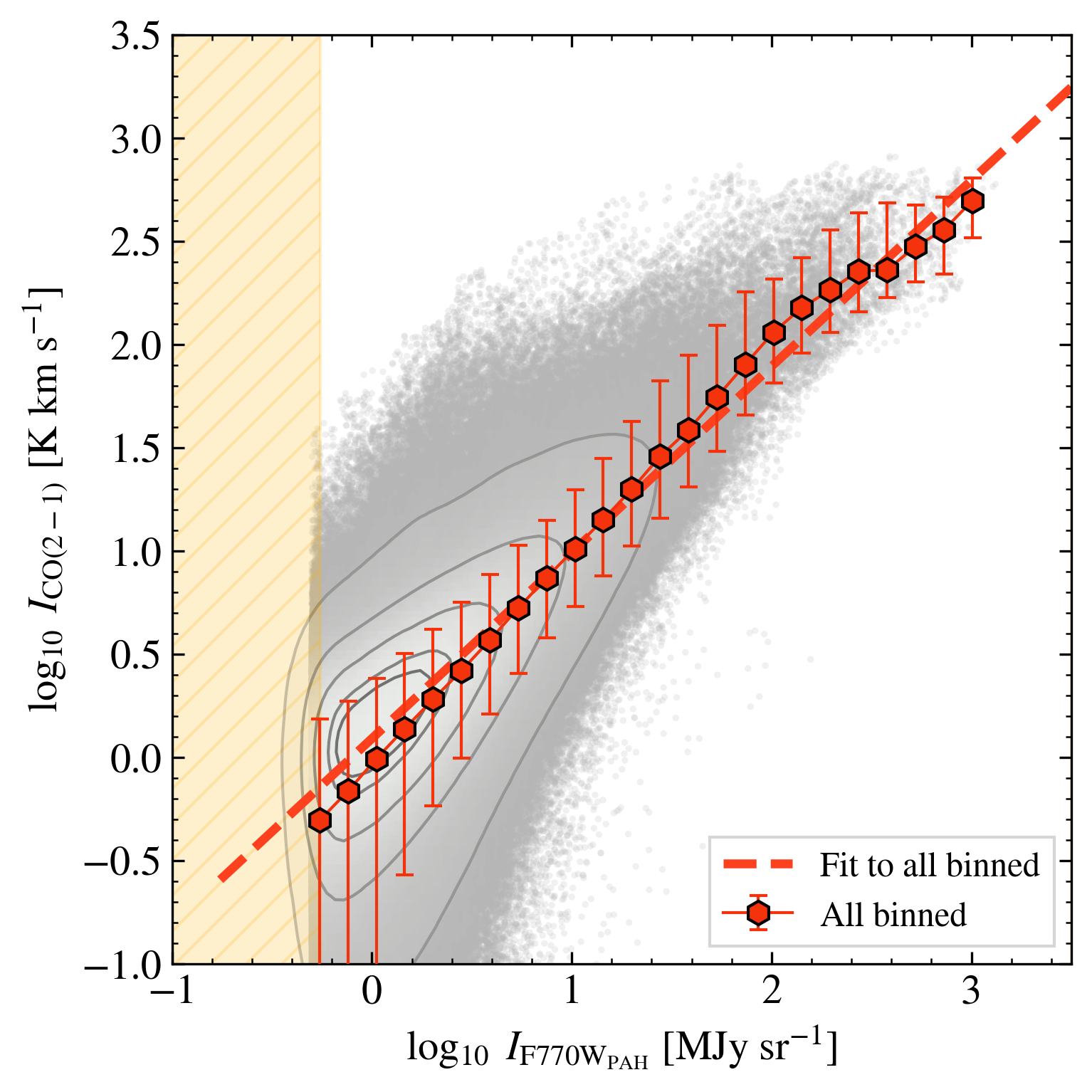}\includegraphics[width=0.5\textwidth]{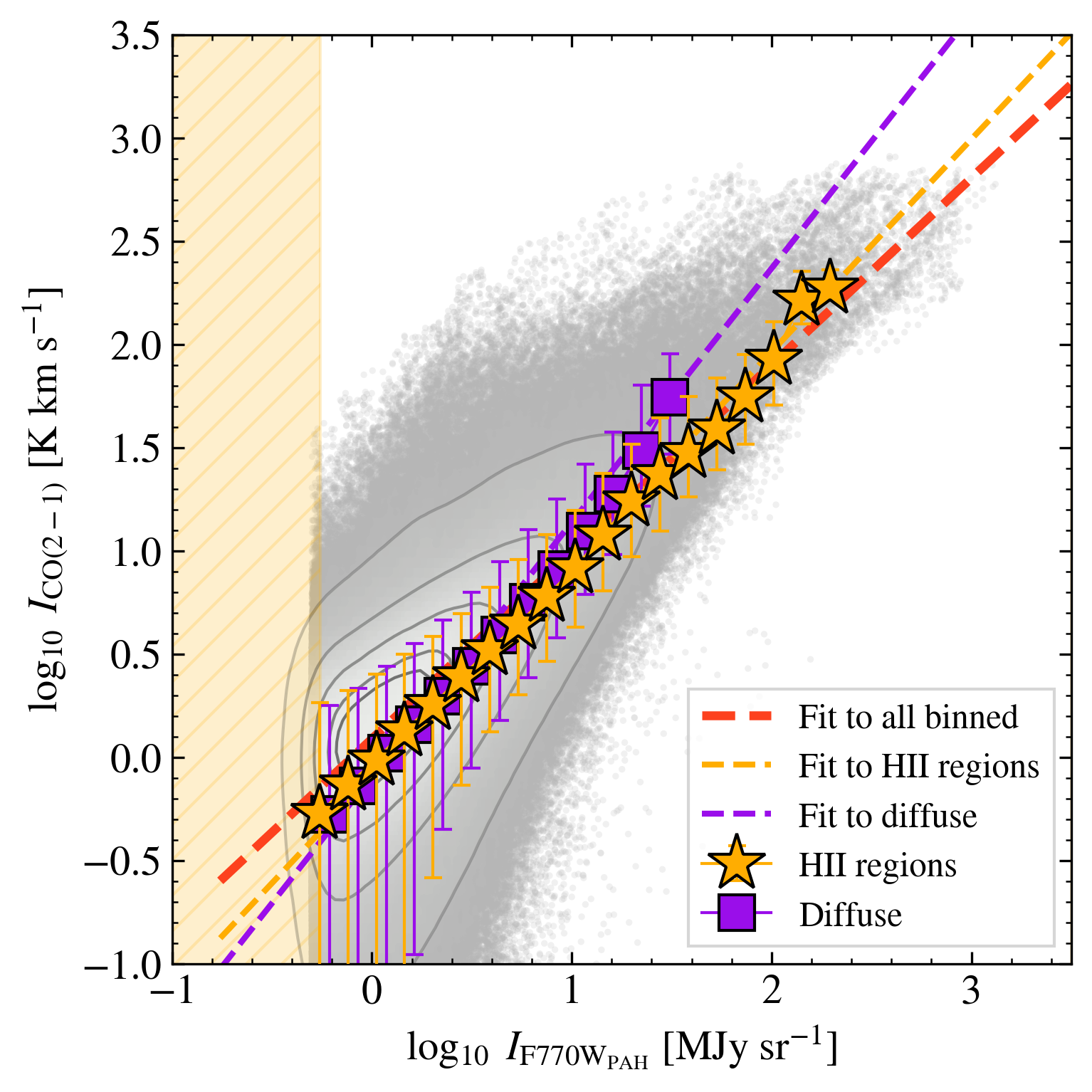}
\includegraphics[width=0.5\textwidth]{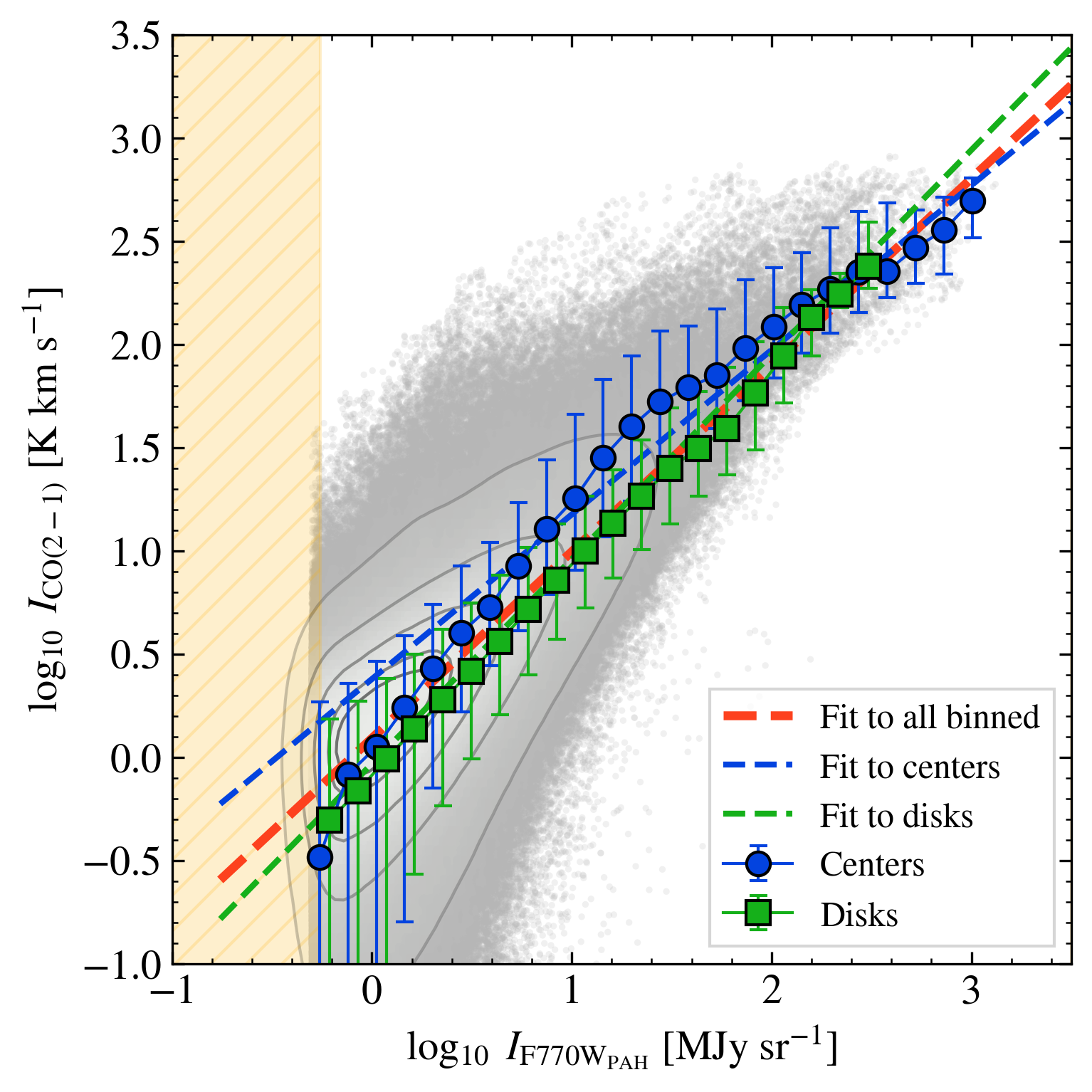}\includegraphics[width=0.5\textwidth]{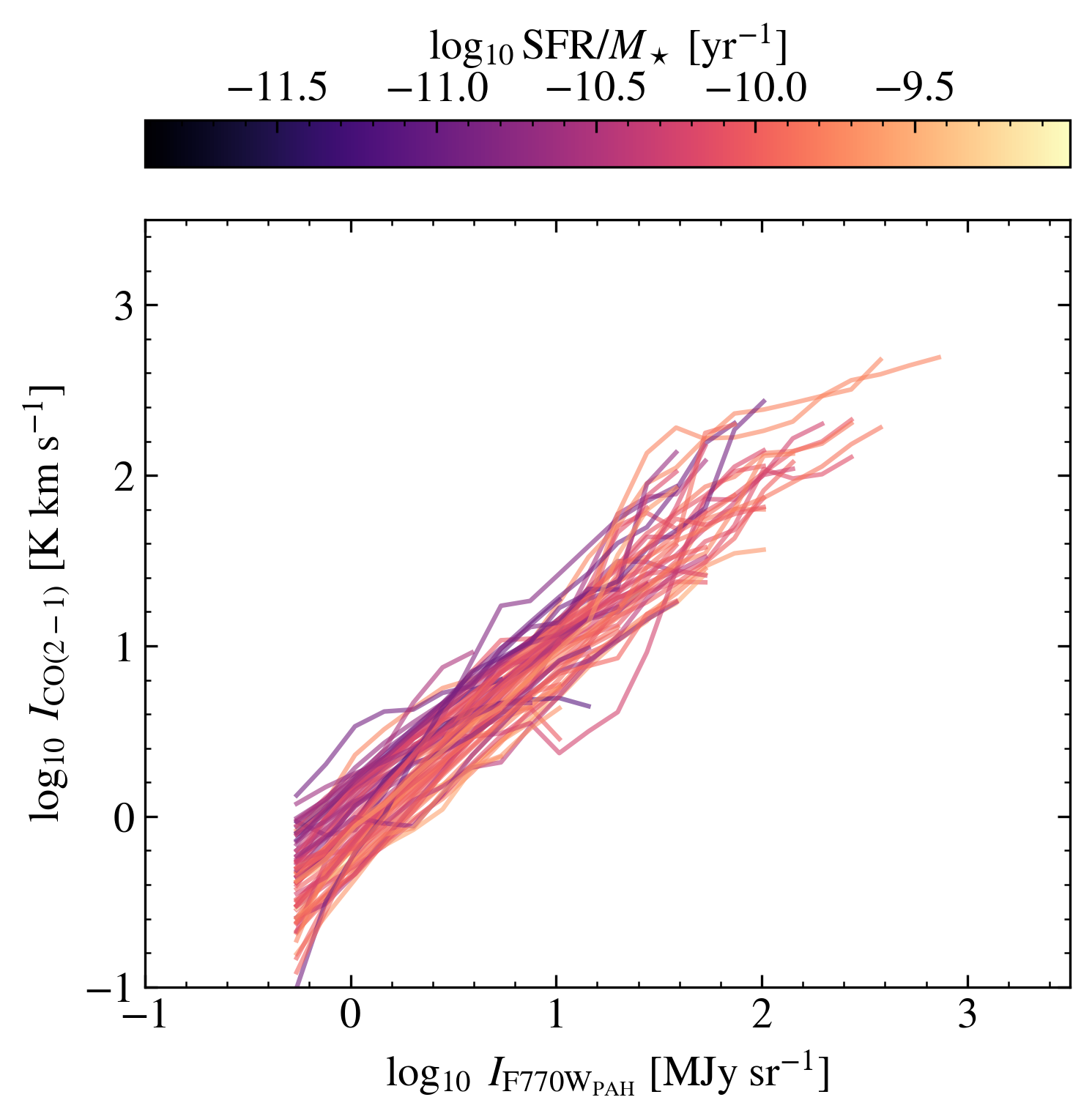}
\caption{\textbf{CO(2--1) and starlight continuum-subtracted F770W$_{\rm PAH}$ emission} at $\approx 100$~pc resolution in 70 nearby star-forming galaxies. \textit{Top left:} All sight lines in our analysis (gray points) with data density contours enclosing the densest 15, 25, 50, 75, and 95\% of the data points. The bins show the median and 16{-}84\% range of CO(2-1) emission in logarithmically-spaced bins of $I_{\rm PAH}^{\rm F770W}$; treating the PAH emission as the independent variable allows us to average the noisier CO(2-1) data. The dashed line shows the best-fit power law describing these binned measurements (Table \ref{tab:fits}).  \textit{Top right:} As the top left panel but now separately plotting results for sight lines near \ion{H}{2} regions (yellow, stars) and diffuse emission outside these regions (purple, squares). The two environments show similar CO-to-PAH ratios where they overlap, but the sight lines near \ion{H}{2} regions show overall brighter intensities. \textit{Bottom left:} As for the previous figures, but now separating galaxy centers (blue, circles) from disks (green, squares). Galaxy centers show brighter emission and higher CO-to-PAH ratios at the same $I_{\rm F770W}^{\rm PAH}$. \textit{Bottom right:} Traces show binned results for each individual galaxy. The galaxies show overall similar CO(2-1) vs.\ F770W$_{\rm PAH}$ relations with moderate offsets from galaxy to galaxy. These offsets correlate with the integrated galaxy properties (see Figure~\ref{fig:co_mir_bygal}, with the color bar indicating SFR$/M_\star$). See Table~\ref{tab:fits} for ratios and best fits for each panel.
\label{fig:co_mir_all}}
\end{center}
\end{figure*}

The top left panel of Figure~\ref{fig:co_mir_all} shows the correlation between CO~(2-1) and star-subtracted F770W, \ipah, for all F770W-bright pixels (\S \ref{subsec:approach}) in all galaxies in our sample. Combining all 70 galaxies we find a strong correlation between CO~(2-1) and starlight-subtracted F770W emission at $50{-}150$~pc resolution, with $r\approx 0.64$. Table~\ref{tab:fits} reports the best-fit relation derived from fitting the binned CO~(2-1) as a function of \ipah. Our measurements of the normalization $I_{\rm CO(2-1)} / I_{\rm F770W}^{\rm PAH} \approx 1.1$~K~km~s$^{-1}$\,(MJy~sr$^{-1}$)$^{-1}$ and the slope (Eq.~\ref{eq:line}) $m=0.90\pm 0.07$, agree reasonably well with previous work on much smaller samples or at lower resolution \citep{chown2021,leroy2023a,leroy2023a}. 

Our selection of bright pixels aims to include regions dominated by molecular gas. However, in the lowest \ipah\ bins ($1\lesssim \ipah \lesssim 30$~MJy~sr$^{-1}$), we do see evidence of a slightly steeper relation, indicating lower CO-to-PAH ratios. This likely indicates a contribution of PAH emission associated with atomic gas (or perhaps CO-dark H$_2$) in these bins. At $\ipah \lesssim 0.1$~\mjysr, we would expect this to become the dominant effect because \ion{H}{1}-dominated regions feature little-to-no CO emission while PAHs may remain present and excited there \citep{boulanger1988}. In the aforementioned low intensity regime of Figure~\ref{fig:co_mir_all}, this would lead to a steeper, more scattered CO-vs-PAH relation, and likely a stronger correlation between \ion{H}{1} column density and \ipah. %

At higher \ipah\ the correspondence between CO and PAH emission appears stronger, with only modest $\approx \pm 0.2$~dex scatter in the CO-to-PAH ratio in the high intensity ($\approx 100$~\mjysr) bins. At lower \ipah, which corresponds to most of the area, the scatter in CO at fixed \ipah\ appears somewhat higher, $\pm 0.5$~dex. Much of this reflects the statistical noise in the PHANGS--ALMA CO maps, which often approaches $\approx \pm 1$~K km~s$^{-1}$ in the ``flat'' CO maps that we use. Our averaging approach recovers the median trend well, but this leads to a large scatter in the data in faint regions of each galaxy. We note that the noise is normally-distributed, which leads to asymmetric scatter when shown in log-space.

\subsection{Nebular and diffuse regions}
\label{sec:nebular}

A number of studies have already demonstrated suppression of PAH emission, likely due to PAH destruction, within \hplus\ regions \citep[e.g.,][]{madden2006, povich2007, gordon2008, montillaud2013,chastenet2023,egorov2023,pedrini2024,sutter2024}, while the intense radiation fields around \hplus\ regions also lead them to stand out as the brightest features in MIR maps of galaxy disks \citep{pathak2024}. We note that at the physical resolution of PHANGS/JWST, PAH emission is reduced but still detected inside \hplus\ regions.  %

The top-right panel in Figure~\ref{fig:co_mir_all} separates sight lines in galaxy disks into those near \hplus\ regions and diffuse (i.e., all other) regions, and Table~\ref{tab:fits} presents our correlation analysis for these two regions separately. Perhaps surprisingly, we find median CO/F770W$_\mathrm{PAH}$ ratios in \hplus\ and diffuse regions to be quite similar, with \hplus\ regions exhibiting a ratio only $\approx 0.10$~dex ($1.25\times$) lower than diffuse regions (i.e., PAHs are slightly brighter relatively to CO near \hplus\ regions). The correlation strength near \hplus\ regions appears stronger than in diffuse regions ($r \approx 0.8$ vs. $r \approx 0.5$), but this partially reflects that the sight lines near \hplus\ regions tend to be brighter, and so less affected by the high statistical noise in the CO maps. The slightly steeper relationship observed for the diffuse regions reflects that diffuse sight lines with \ipah\ $\gtrsim 10$~\mjysr\ show slightly elevated CO/PAH ratios compared to nebular regions with similar intensities, which may be due to preferential destruction of CO compared to PAHs in nebular regions. This preferential destruction may be related to recent work demonstrating that PAH emission is slightly more long-lived than CO emission in gas-poor regions (Kim et al. in preparation). Sight lines with \ipah\ $\lesssim 10$~\mjysr\ in Figure~\ref{fig:co_mir_all} show almost identical CO-to-PAH ratios for diffuse and nebular regions. 

The similarity of the CO-vs-PAH relationship in diffuse and nebular regions likely results from the fact that the \hplus\ region masks that we use \citep[from][]{groves2023} are based on data with coarse resolution (median $70$~pc) compared to the actual size of most \hplus\ regions \citep[e.g., see][]{barnes2022}. As a result, these regions often include CO and PAH emission from ISM material projected towards, but not actually inside \hplus\ regions. The CO and PAH emission may come from the well-shielded material surrounding the \hplus\ regions. \citet{sutter2024} discuss a scenario in which the \citet{groves2023} regions contain a mixture of diffuse material and true \hplus\ regions to explain observed F770W/F2100W ratios. Resolved comparisons of high ($\lesssim 10$~pc) physical resolution H$\alpha$, Pa$\alpha$, CO, and MIR emission \citep[e.g.,][]{pedrini2024} should help test this hypothesis in the near future. Subtracting hot dust continuum emission from F770W, which we do not do, may also be important, since the filter still captures such emission in regions where PAHs have been destroyed to the point where their emission is not detectable.

\subsection{Centers and disks of galaxies}
\label{sec:centers}

Galaxy centers, especially bar-fed central molecular zones, differ from the disks of galaxies in ways that may also affect CO-to-PAH ratios. Galaxy centers exhibit some of the most intense star formation found in galaxies, with correspondingly high interstellar radiation fields, gas column and volume densities, and some host active galactic nuclei \citep[e.g.,][and see review in \citealt{schinnerer2024}]{schinnerer2023}. As a result, CO in galaxy centers often exhibits broader line widths, lower opacity, and low $X_{\rm CO}$ \citep[e.g.,][]{bolatto2013,teng2023}.

The bottom-left panel of Figure~\ref{fig:co_mir_all} shows the CO-vs-PAH relationship separating sight lines towards galaxy centers compared to those in disks. Both CO and PAH emission appear systematically brighter in galaxy centers than in disks, likely due to the higher densities and stronger radiation fields in galaxy centers. This is expected from previous analyses of these galaxies \citep[e.g.,][]{sun2020,pathak2024}. Sight lines in galaxy centers also appear offset towards higher CO intensity compared to disks at matched \ipah. On average the CO-to-PAH ratio is  $\approx 0.16$~dex ($\approx 60\%$) higher in galaxy centers compared to disks, and the median $I_{\rm CO(2-1)}$ in centers appear higher than that found for disks at fixed \ipah.

We note one caveat here. The stellar continuum can also be bright in galaxy centers, and the starlight subtraction can become correspondingly more difficult \citep[e.g.,][]{sutter2024,baron2024}. If our standard correction oversubtracts the starlight from F770W in galaxy centers, then we might artificially underestimate F770W$_\mathrm{PAH}$. However, considering a similar situation, \citet{baron2024} found that the range of plausible starlight SED variations cannot impact the F770W emission enough to explain observed variations in the F770W/F1130W ratio in the inner parts of galaxies. We also note that we find a similar contrast between disks and centers in the CO-to-F1130W ratio, where F1130W is less affected by starlight. Finally, we note that we do not subtract any hot dust continuum from F770W. Hot dust continuum is likely stronger in galaxy centers, implying that our reported CO-to-PAH ratio measurements are lower than the true ratios.

A higher CO-to-PAH ratio in bright galactic centers may imply a large impact of $X_{\rm CO}$ on the measured ratio. A number of studies have shown that $X_{\rm CO}$ in galaxy centers can be 5--15 times lower than the standard Milky Way $X_{\rm CO} = 2\times 10^{20}$~cm$^{-2}$/(K~km~s$^{-1}$), and $\approx 2\times$ lower than the galaxy mean on average \citep[e.g., ][]{sandstrom2013,israel2020,teng2023,chiang2024}. Following Equation \ref{eq:dust}, centrally suppressed $X_\mathrm{CO}$ will lead to centrally enhanced CO-to-PAH ratios. On the other hand, the intense radiation fields expected in galaxy centers should increase PAH emission and decrease the CO-to-PAH ratio, unless PAH destruction occurs. There is some evidence for slightly lower \qpah\ in galaxy centers \citep{chastenet2023a}, but a general trend is not so clear. Metallicity and DGR tend to be highest in galaxy centers, which may also decrease the CO-to-PAH ratio (see Eq.~\ref{eq:dust}). Considering that most of these effects would decrease CO-to-PAH, the enhancement seen in Figure~\ref{fig:co_mir_all} and Table \ref{tab:fits} may indicate that $X_{\rm CO}$ represents the dominant effect, offsetting these other PAH-enhancing effects. %

\subsection{Galaxy to galaxy variations}
\label{sec:galtogal}

\begin{figure*}
\begin{center}
\includegraphics[width=0.5\textwidth]{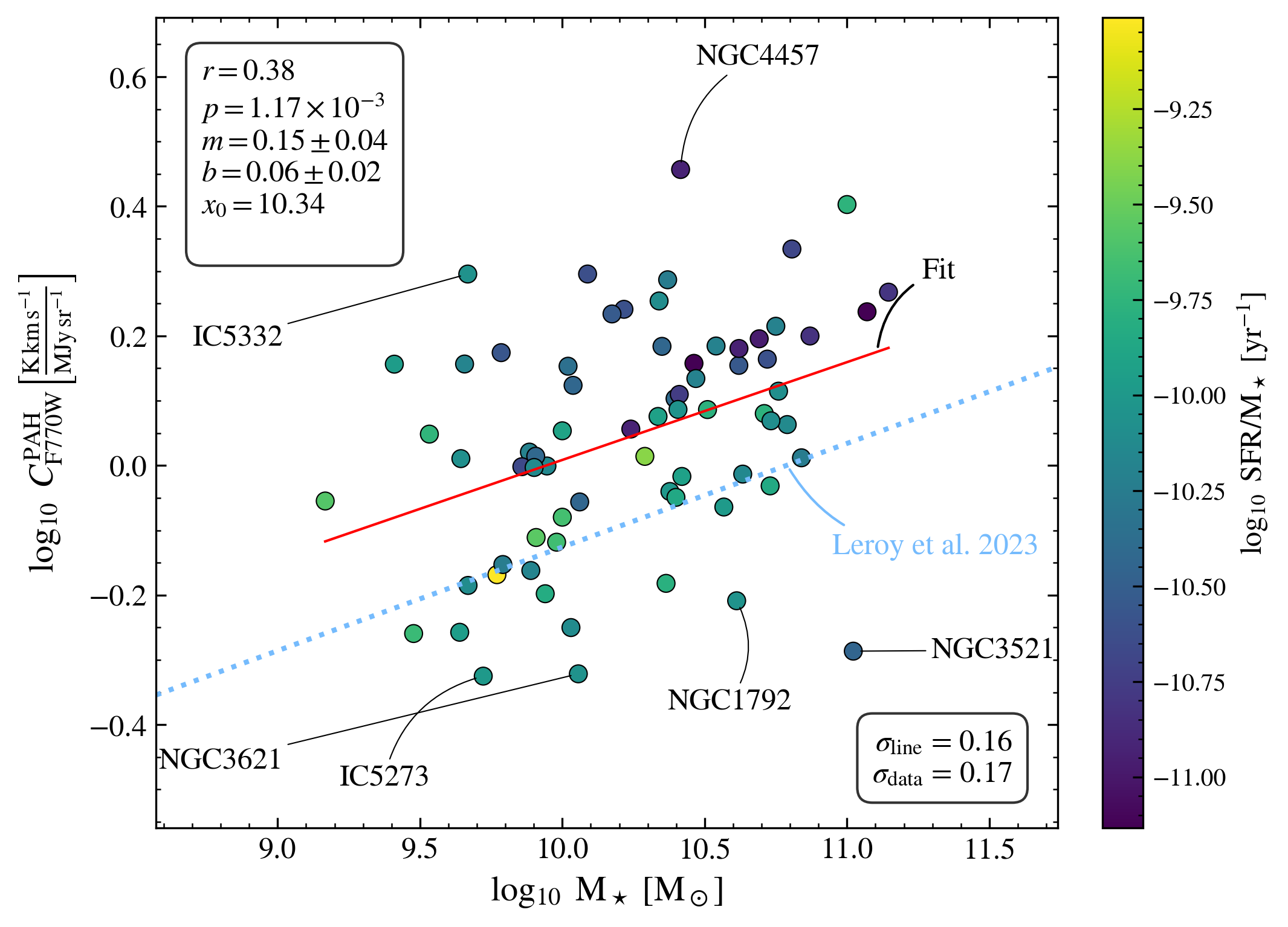}\includegraphics[width=0.5\textwidth]{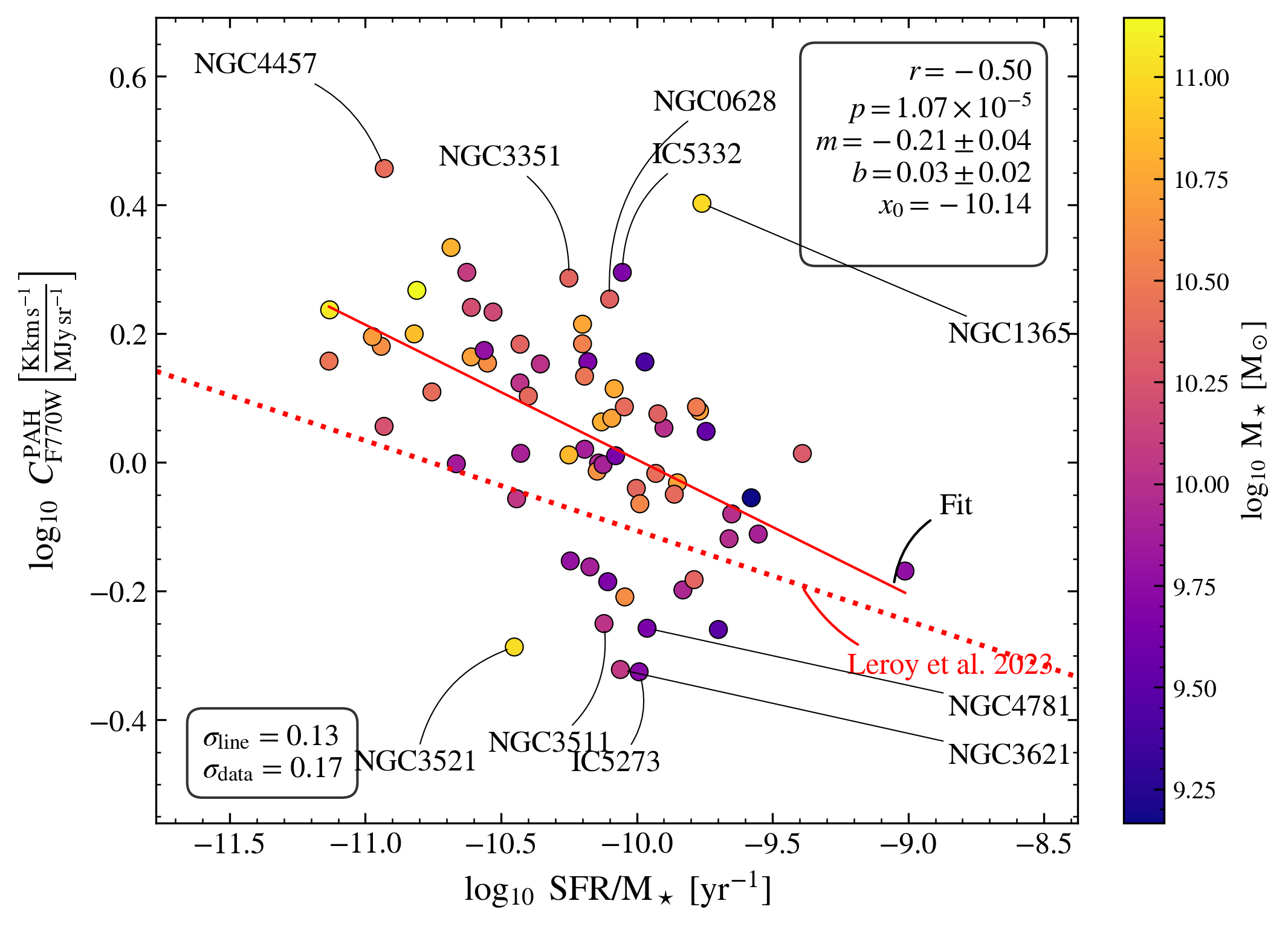}
\includegraphics[width=0.5\textwidth]{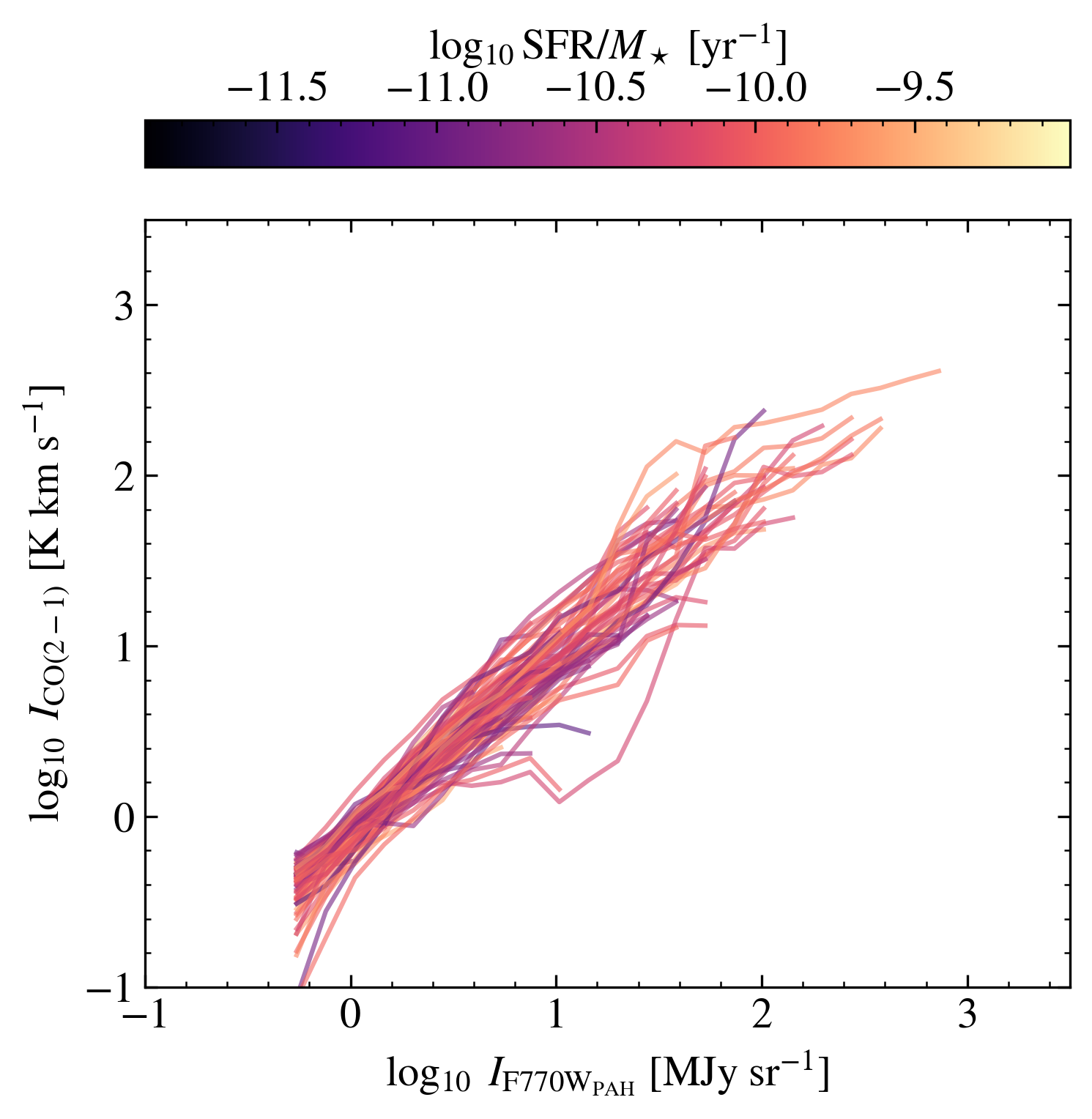}\includegraphics[width=0.5\textwidth]
{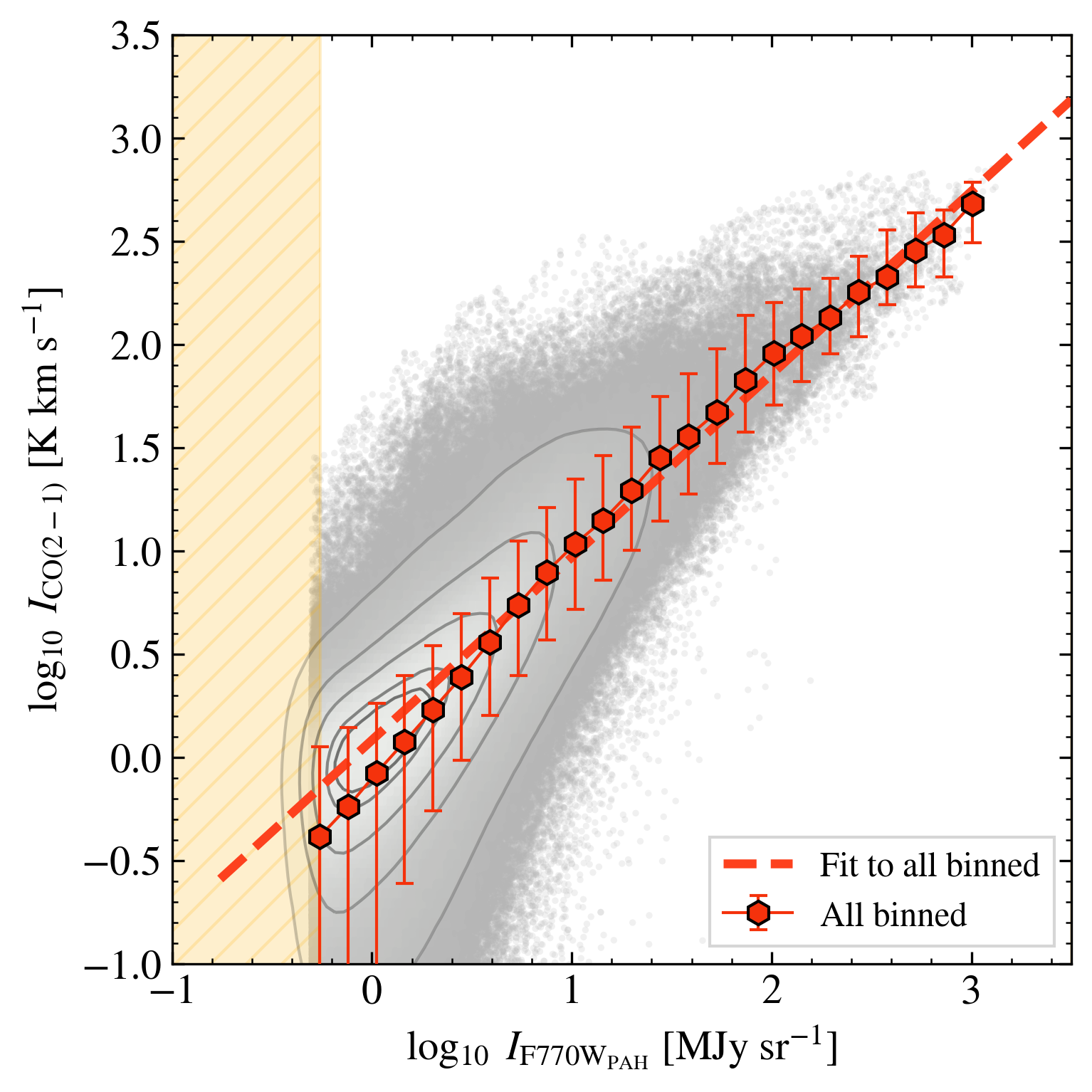}
\caption{\textbf{Galaxy-to-galaxy variations in the CO(2--1) vs. F770W$_{\rm PAH}$ relationship.} CO(2-1) at fixed PAH intensity for individual galaxies (see bottom left panel Fig.~\ref{fig:co_mir_all}) as a function of (\textit{left}) galaxy-integrated stellar mass, $M_\star$ and (\textit{right}) specific star formation rate, SFR/$M_\star$. Spearman's $r$, p-value, the best-fit slope, intercept and pivot ($y=m(x-x_0)+b$) to all galaxies, the scatter about the best-fit relation $\sigma_\mathrm{line}$ (Eq.~\ref{eq:norms}), and vertical scatter in the normalization $\sigma_\mathrm{data}$ are all indicated.  Scaled versions of the best-fit normalizations of $I_\mathrm{CO(2-1)}$ vs \textit{WISE} 12~\um\ against stellar mass and SFR$/M_\star$ from \citet{leroy2023a} are shown (dotted lines), showing similar trends. Galaxies further than $1.5\sigma_\mathrm{line}$ from the best-fit lines are labeled. We observe a modest correlation between the CO-to-PAH ratio and $M_\star$ and a well-defined anti-correlation between the CO-to-PAH ratio and SFR/$M_\star$. The correlation with $M_\star$ may reflect increased contribution of PAH emission associated with CO-dark gas or atomic gas in low mass galaxies. The anti-correlation with SFR/$M_\star$ likely reflects a mixture of increased radiation field strength, lower $X_{\rm CO}$, and enhancement of $R_{21}$ in high SFR/M$_\star$ galaxies, which appear to represent stronger effects than any $q_{\rm PAH}$. The bottom left panel shows the CO-vs-F770W$_{\rm PAH}$ relationships for each galaxy normalized by $C^{\rm PAH}_{\rm F770W}$ predicted based on its $\log_{10}$SFR/$M_\star$. This normalization reduces the vertical scatter. The bottom right panel shows the fit to all pixels and all galaxies with the normalizations applied. We provide fit parameters for F770W$_{\rm PAH}$ in Eq.~\ref{eq:copah_renorm} and for the rest of the bands in Table~\ref{tab:norm_fits}.
\label{fig:co_mir_bygal}}
\end{center}
\end{figure*}

So far, we have treated all galaxies together, separating sight lines into categories but not otherwise differentiating between targets. However, our sample spans a range of stellar mass (\mstar), star formation rate (SFR), metallicity, morphology, and more. These factors will influence $U$, DGR, \qpah, and $X_{\rm CO}$ and so we might expect differences among the CO-PAH relationships for different galaxies \citep[as in][]{chown2021}. To explore the impact of these variations, the bottom-right panel of Figure~\ref{fig:co_mir_all} shows the CO-to-F770W relationship varies from galaxy to galaxy. The fits for individual galaxies are provided in Appendix~\ref{appendix:allgals}.

The bottom right panel of Figure~\ref{fig:co_mir_all} shows that individual galaxies exhibit strong CO-PAH correlations, mostly parallel to our best-fit overall relation. Broadly, the agreement among the 70 individual galaxies appears good, supporting the potential use of PAHs to trace CO. Specifically, the normalization of the CO-PAH relation scatters by $\pm 0.17$~dex from galaxy to galaxy. Simply applying our best fit overall relation to an individual galaxy with no additional information can be expected to yield a map biased by a factor drawn from this galaxy-to-galaxy scatter. This is similar to the pixel by pixel scatter observed \textit{within} each galaxy, $\sigma_\mathrm{pix}$ in Table~\ref{tab:galtab}, implying that galaxy-to-galaxy variations and internal scatter contribute about equally to the total observed scatter.

In Figure~\ref{fig:co_mir_bygal}, we test how these galaxy-to-galaxy offsets correlate with integrated galaxy properties. We compute the normalization of the best-fit CO(2-1) versus F770W$_\mathrm{PAH}$ power law for each galaxy, \norm, by performing a linear fit of $\log_{10}I_\mathrm{CO(2-1)}$ versus $\log_{10}\ipah$ for each galaxy. \norm\ is then given by the value of the best-fit relation at $\ipah = 1$~MJy~sr$^{-1}$. We plot \norm\ against \mstar\ and specific star formation rate (\ssfr) \citep[WISE+GALEX-based galaxy-integrated SFR and $M_\star$ are drawn from][]{leroy2021}. Stellar mass correlates with the H$_2$/\ion{H}{1} ratio, gas-phase metallicity, DGR, $X_{\rm CO}$ and more \citep[e.g.,][]{saintonge2022}. Meanwhile, \ssfr\ anti-correlates with \qpah, and correlates with the mean interstellar radiation field, $\bar{U}$ \citep{chastenet2024}. Therefore following Equation \ref{eq:dust} both parameters might be expected to impact the CO-to-PAH ratio. Correlations between the galaxy-integrated CO-to-12$\mu$m ratio (with the $12\mu$m data from WISE) with star formation activity, stellar mass, and \ssfr\ have previously been observed \citep{chown2021,leroy2023a}.

Consistent with previous work we find a mild positive correlation between the CO/PAH ratio at fixed \ipah\ and \mstar\ and an anti-correlation between the CO/PAH normalization and \ssfr. As seen in the bottom right of Figure~\ref{fig:co_mir_all}, the galaxy-to-galaxy offsets are thus not random but agree with physical expectations. The $\log_{10}$\norm\ vs. $\log_{10}$\ssfr\ trend may indicate the higher $U$ associated with high \ssfr\ galaxies leads to stronger PAH emission, offsetting any suppression due to lower $q_{\rm PAH}$ at high \ssfr . As mentioned in \S\ref{sec:motivation}, the anti-correlation between \norm\ and \ssfr\ is expected because the dust heating rate increases with \ssfr. 
The $\log_{10}$\norm\ vs $\log_{10}M_\star$ trend goes in the sense that galaxies with higher metallicity, DGR, and molecular-to-atomic gas ratios \citep[e.g.,][]{saintonge2017} show higher CO-to-PAH ratios. As a result, perhaps the low observed CO-to-PAH ratios seen in low $M_\star$ systems reflects that the PAH emission is coming from regions dominated by \hone\ or CO-dark H$_2$ despite our selection of only bright emission.

We also checked for correlations with distance and inclination to test how orientation and resolution might bias our results. We found that the normalization is uncorrelated with distance ($r=-0.04, p=0.78$), and weakly correlated with $\cos i$ ($r=0.24, p=0.06$). Variations in $\cos i$ from galaxy to galaxy simply slide the data points parallel to a 1:1 relation. Since the best-fit slopes of CO(2-1) vs. PAH emission for each galaxy and for the sample as a whole are within a few percent of 1.0, and furthermore that the sample covers a range of inclinations across the full $M_\star$ range, it is understandable that $\cos i$ does not have a significant effect.

We fit functional forms to the two trends and note the \ssfr\ as the stronger, clearer trend. These predict the normalization of the CO vs. PAH relation as functions of galaxy-integrated $M_\star$ or \ssfr 
\begin{eqnarray}\label{eq:norms}
\nonumber \log_{10}\norm = 0.15 \pm 0.04~(\log_{10} M_\star -10.34) \\
 +0.06 \pm 0.02, \\
\nonumber \log_{10}\norm = -0.21 \pm 0.04~(\log_{10}\ssfr +10.14)\\
 +0.03 \pm 0.02.
\end{eqnarray}
We provide best-fit parameters for normalizations versus $\log_{10}M_\star$ and $\log_{10}\mathrm{SFR}/M_\star$ for the other PAH bands in Table~\ref{tab:norm_fits}.

In the bottom panels of Figure~\ref{fig:co_mir_bygal} we normalize the data from each galaxy by the values predicted by this fit, aiming to remove galaxy-to-galaxy scatter. Then we re-fit the relation to all data (lower right panel) and find
\begin{equation}\label{eq:copah_renorm}
\log_{10}I_\mathrm{CO(2-1)}=(0.88\pm 0.06)(x-1.44)+(1.36\pm 0.06), 
\end{equation}
with scatter $\sigma=0.43$~dex, where $x\equiv \log_{10}\ipah - \log_{10}\norm$ shown in Table~\ref{tab:big_fit_table_normed} along with prescriptions for the other two PAH bands. We  describe how to use these prescriptions in Appendix~\ref{appendix:howto}.

\subsection{CO~(2-1) and PAH emission for other bands}
\label{sec:co_pah_bands}

\begin{figure*}
\begin{center}
\includegraphics[width=0.48\textwidth]{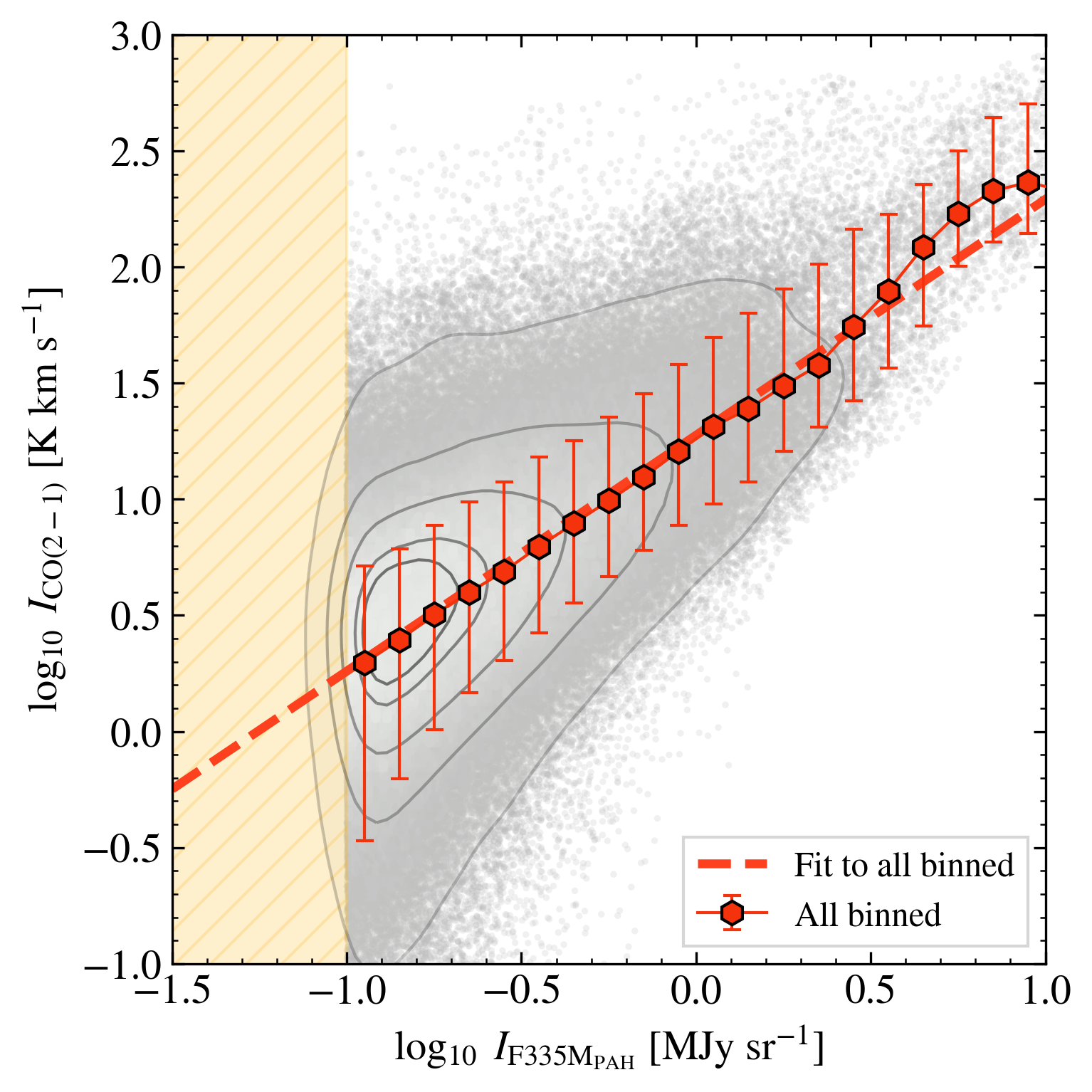}
\includegraphics[width=0.48\textwidth]{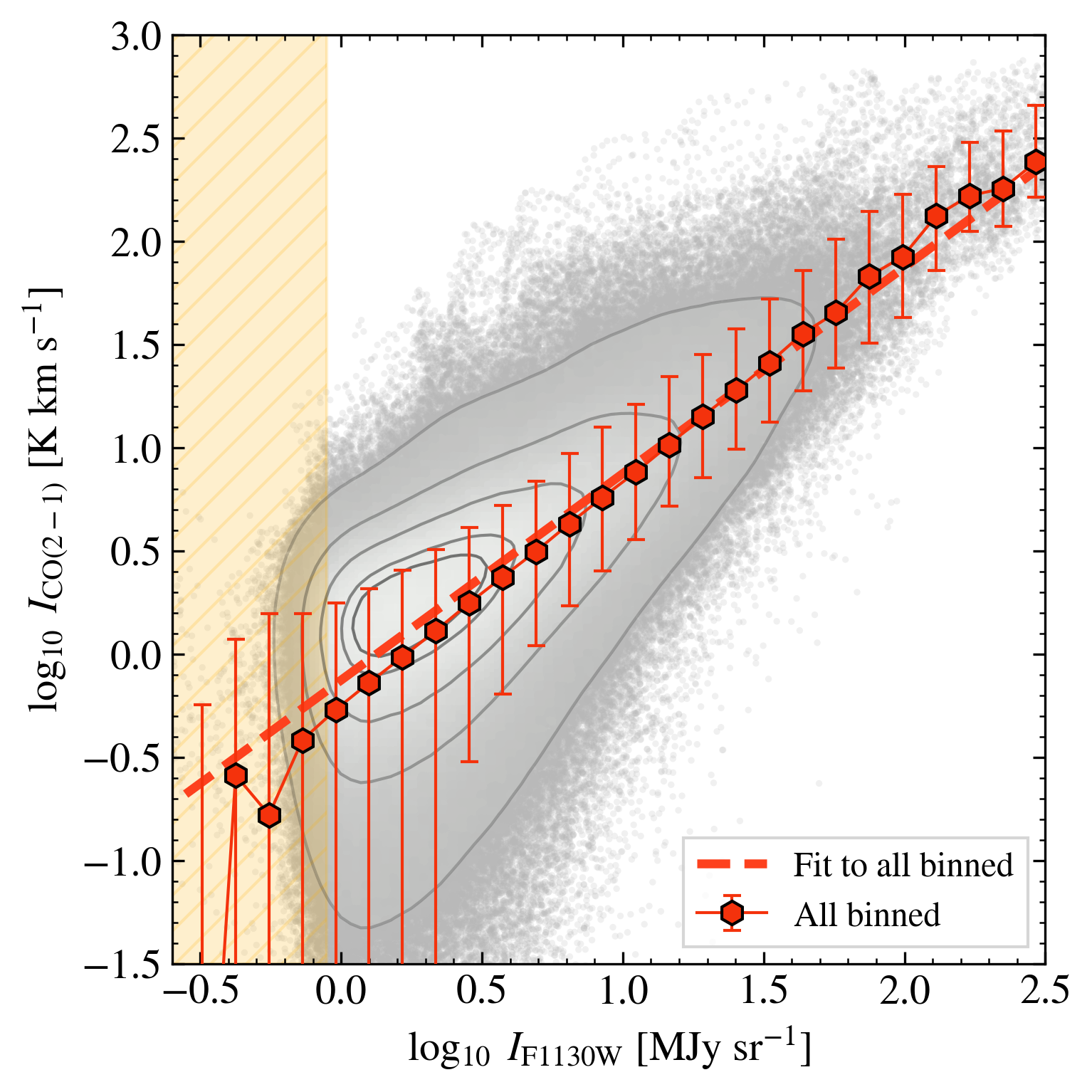}
\caption{\textbf{CO(2--1) intensity as functions of 3.3~$\mu$m and $11.3~\mu$m PAH intensity.} As Figure~\ref{fig:co_mir_all}, but now showing PAH intensity captured by (\textit{left}) the F335M filter capturing the 3.3~$\mu$m PAH feature (after continuum subtraction following H. Koziol et al. in preparation) and (\textit{right}) the F1130W filter capturing the 11.3~$\mu$m PAH feature. The relationship between CO(2-1) and these bands resembles that which we observe for F770W$_{\rm PAH}$ in Figure~\ref{fig:co_mir_all}, though the specifics of the fits differ (Table~\ref{tab:fits}). Of note, the F335M$_{\rm PAH}$ feature has a lower intensity than the others and extracting it from imaging depends critically on stellar continuum subtraction, but the shorter wavelength means that the band offers even higher resolution compared to the other features \citep[see][]{sandstrom2023}. %
\label{fig:co_mir_all_otherbands}}
\end{center}
\end{figure*}

So far we have focused only on the correlation between CO(2-1) and F770W$_{\rm PAH}$. Do the other PAH-tracing filters show similar correlations? In Figure~\ref{fig:co_mir_all_otherbands} we show CO versus F335M$_\mathrm{PAH}$ and versus F1130W for our $19$ Cycle 1 targets. The median CO/PAH ratios, fits and statistics for these bands are shown in Table~\ref{tab:fits}. The correlation remains strong in both of these bands, and the best fit binned CO-PAH slopes are also very close to linear for the $3.3~\mu$m and $11.3~\mu$m features. Of the three PAH bands, F335M$_\mathrm{PAH}$ is the faintest \citep[][]{chastenet2023a,sandstrom2023}, and therefore shows the largest CO-to-PAH ratios ($10^{1.24} \approx 17.4$ times higher at F335M$_\mathrm{PAH}$ than F770W$_\mathrm{PAH}$; while F1130W is a factor of $10^{0.15} \approx 1.4$ times lower than F770W$_\mathrm{PAH}$); see top rows of Table~\ref{tab:fits}.

These three PAH bands (3.3, 7.7, and 11.3~\um) are dominated by different species of the PAH population -- smaller, neutral PAHs for 3.3~\um, ionized PAHs for a range of sizes for 7.7~\um, and larger, mainly neutral PAHs for 11.3~\um. Their intensity ratios also respond to changes in the interstellar radiation field spectrum \citep[e.g.,][]{maragkoudakis2020,draine2021}. All three bands yield reasonable first-order estimators of CO intensity, but we would not expect all three to show identical or even equally good correlations with CO. We anticipate that future work will explore optimal combinations of bands to trace CO (and H$_2$) and examine the impact of conditions in the molecular gas on PAH band ratios.

From a practical perspective, each band has advantages. The F770W band traces the brightest PAH feature and is available now for $>70$ nearby galaxies. The F335M filter offers even sharper resolution but harbors the fainter 3.4~\um\ PAH feature, the Pfund-$\delta$ line \citep[e.g.][]{peeters2024}, and is more strongly affected by contamination by starlight. Meanwhile, the F1130W filter is largely unaffected by starlight and also captures a bright feature in a relatively narrow filter.

\section{Discussion}
\label{sec:discussion}

\begin{figure*}
\begin{center}
\includegraphics[width=\textwidth]{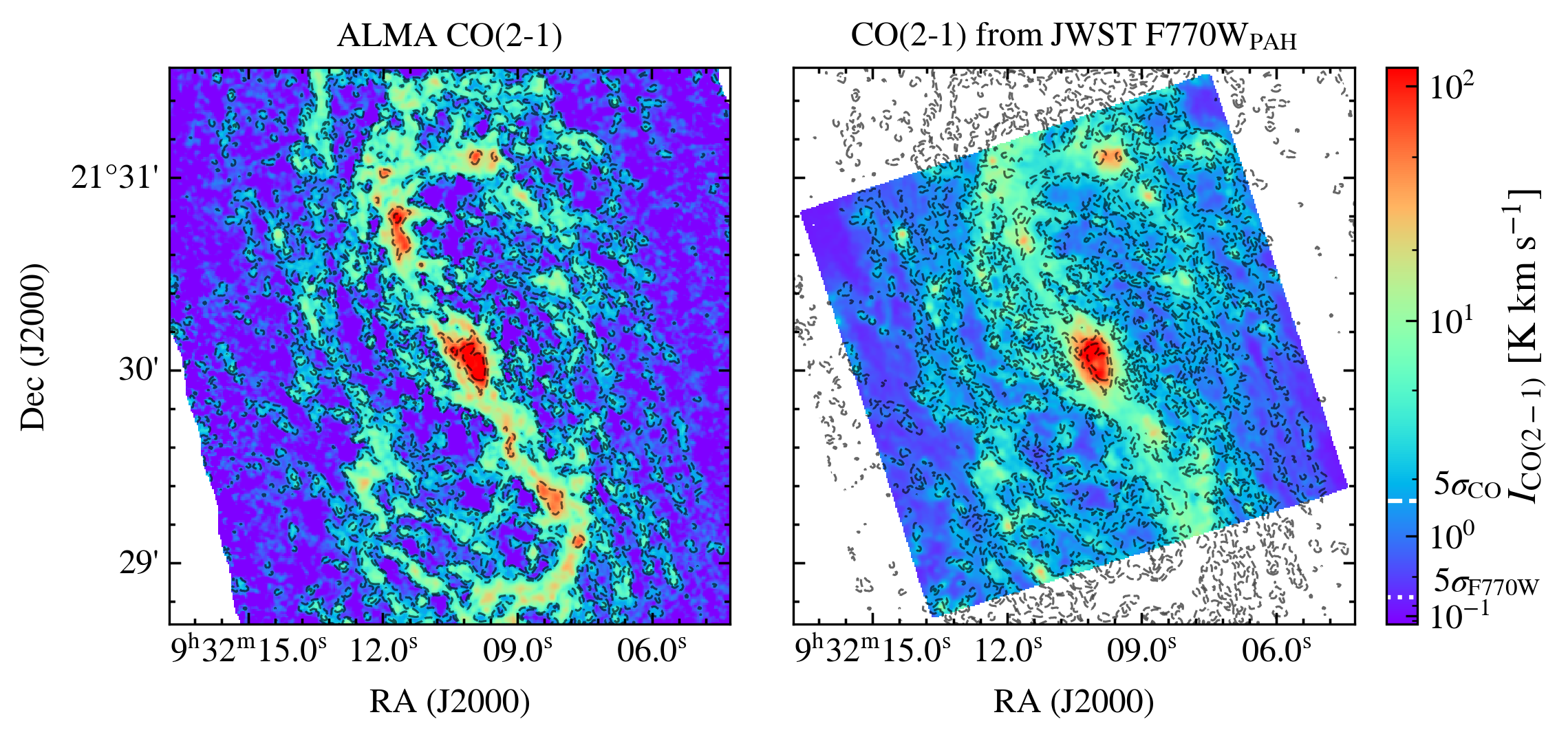}
\caption{\textbf{Predicting CO from PAH emission in one nearby galaxy.} The left panel shows an inclination-corrected ALMA CO(2-1) image of NGC~2903 with contours at 1.25, 3.1, 31, and 100 K~km~s$^{-1}$. The right panel shows a predicted CO(2-1) intensity map based on JWST F770W$_\mathrm{PAH}$ and the prescription in Eq.~\ref{eq:copah_renorm}. The F770W$_\mathrm{PAH}$ image is masked to the NIRCam footprint as required for starlight subtraction (\S\ref{subsec:jwst}). The contours, which are the same in both panels, show a very close correspondence between CO and PAH emission. At the same time, CO(2-1) emission is underestimated in the bar ends, highlighting improvements to be explored in future work. This figure also shows the improved sensitivity of JWST compared to ALMA at recovering faint emission, with the $5\sigma$ RMS noise indicated on the colorbar. 
\label{fig:demo}}
\end{center}
\end{figure*}

We show that PAH emission can be used to predict CO(2-1) emission with $\approx 0.5$~dex (i.e., a factor of $\approx 3$) scatter at $100$~pc resolution in the disks of star-forming galaxies, without requiring any other information. Doing so, one expects an $\approx \pm 0.2$~dex bias in any given prediction due to galaxy-to-galaxy variations in the CO-to-PAH ratio, which correlates with both $\log_{10}\ssfr$ and $\log_{10}\mstar$. We provide prescriptions to account for these galaxy-to-galaxy variations, which can sharpen the prediction even more. Fig. \ref{fig:demo} shows an example of such a prediction for the nearby spiral NGC~2903.

Our results formally apply to regions where molecular gas is likely to constitute a significant fraction of the ISM, with $\Sigma_{\rm mol} \gtrsim 4$~M$_\odot$~pc$^{-2}$. It is likely that in fainter regions the PAH emission reflects the distribution of atomic gas \citep[see][]{sandstrom2023a}, but the details of that correlation remain less well constrained, including the dependence of PAH abundance on \ion{H}{1} phase and density \citep{hensley2022}. We also emphasize that our results offer a way to predict CO emission, specifically CO(2-1) emission for regions with $I^\mathrm{PAH}_{\rm F770W}>0.5$~MJy~sr$^{-1}$. The CO(2-1)-to-H$_2$ conversion factor is known to vary as a function of environment across galaxies and this will need to be included to predict the gas column density, $N({\rm H}_2)$ \citep[e.g., see reviews in][]{bolatto2013,schinnerer2024}. 

Despite these limitations, the CO-PAH correlation represents a powerful tool. In a matter of minutes on the source, JWST can produce maps with resolution and gas column sensitivity that would take ALMA many hours to match (of course ALMA carries kinematic information and CO represents a well-calibrated gas tracer, so the two remain complementary). As an example of the applications of such data, \citet{pathak2024} analyzed $19$ of the same galaxies we study to infer column density probability distribution functions at high physical resolution, suitable for benchmarking simulations and inferring some aspects of interstellar turbulence and galactic dynamics \citep{meidt2023, thilker2023}.

While the practical applications of the observed correlation are exciting, the stability of the CO-to-PAH ratio across a wide range of systems may reflect that the terms in Eq.~\ref{eq:dust} have counter-balancing environmental dependencies. The interstellar radiation field varies within and among galaxies \citep[e.g.,][]{draine2007,aniano2020,chastenet2024}, as does $X_{\rm CO}$, and \qpah\ appears strongly suppressed within \ion{H}{2} regions \citep[][]{chastenet2023,egorov2023,sutter2024}. This might reflect that environmental effects are somewhat offset, e.g., because $U$ is higher where $R_\mathrm{PAH}$ \citep{egorov2023}, as well as $q_{\rm PAH}$ and $\alpha_{\rm CO}$ tend to be low. Alternatively, they might reflect that the emitting PAHs tend to reside predominantly in neutral, moderately shielded gas, subjecting them to some of the same selection effects that apply to CO.

Beyond speculation, having established the basic observational correlations, the clear next step in this area is to follow up with a physically oriented analysis. The full PHANGS data sets make it possible to combine 21-cm data on atomic gas and best-estimate $\alpha_{\rm CO}$ with estimates of the local interstellar radiation field, $U$, and PAH abundance, $q_{\rm PAH}$, to test the physical expectation that $I_{\rm PAH} / N \left( H \right) \propto U q_{\rm PAH}^{-1}$. This will both illuminate the reasons for the tightness of the PAH-CO correlation and sharpen the use of PAH emission as an ISM tracer, particularly in the atomic gas-dominated parts of galaxies.

\section{Conclusions}
\label{sec:conclusions}

We characterize the relationship between CO(2-1) and near- and mid-infrared (NIR and MIR) PAH emission at $\approx 100$~pc scales for 70 nearby ($D \lesssim 20$~Mpc) star-forming galaxies observed as part of PHANGS--ALMA and the PHANGS--JWST Cycle 1 and Cycle 2 treasuries. This is by far the largest comparison of molecular gas tracers and PAH emission at cloud scales to date, more than $10\times$ larger than initial JWST studies. We find:

\begin{enumerate}
\item CO(2-1) exhibits strong correlations with the PAH emission captured by JWST's F770W, F335M, and F1130W filters. In regions of galaxies where molecular gas is likely to make up most of the ISM ($I_{\rm F770W} > 0.5$~MJy~sr$^{-1}$; $\Sigma_{\rm mol} \gtrsim 4$~M$_\odot$~pc$^{-2}$), this correlation appears approximately linear ($0.8 \lesssim m \lesssim 1.2$) and covers more than two orders of magnitude in PAH and CO intensities. We provide power law scaling relations that can be used to predict CO(2-1) from PAH emission (Table \ref{tab:fits}, Figure~\ref{fig:co_mir_all}). The typical sightline-to-sightline scatter about these relations considering the whole sample together is $\sigma \approx 0.5$~dex, and is dominated by statistical noise in the CO measurements in our data set.

\item Subdividing the $19$ JWST Cycle 1 targets into sightlines near \ion{H}{2} regions and diffuse sightlines, we find overall similar scaling relationships between CO(2-1) and $I_{\rm F770}^{\rm PAH}$ between the two types of regions (Table \ref{tab:fits}, Figure~\ref{fig:co_mir_all}). The main difference appears to be that the nebular regions harbor more high-intensity sightlines than the diffuse regions \citep[see also][]{pathak2024}. This is consistent with the idea that the nebular regions still contain some well-shielded, denser gas outside the actual \ion{H}{2} regions that harbor both CO and PAH molecules \citep[see also][]{sutter2024}.

\item We also contrast galaxy centers with emission from the surrounding disks, finding that galaxy centers exhibit on average an $\approx 0.2$~dex ($\approx 60\%$) higher ratio of CO(2-1)-to-F770W emission compared to galaxy disks \citep[in addition to being brighter in both tracers, as in][]{pathak2024}. This might reflect that the enhanced CO emission (i.e., low $\alpha_{\rm CO}$) in galaxy centers \citep[e.g.,][]{teng2023,chiang2024} represents a stronger effect than any enhancement in PAH emission due to a more intense interstellar radiation field. Because of this contrast, when fitting emission from both disk and central regions, the slope of the CO vs.\ PAH relation tends to be somewhat steeper than what we observe for disks or centers alone.

\item Individual galaxies show similar relations between CO(2-1) and PAH intensity, but with $\approx \pm 0.2$ dex scatter in the normalization. The ratio of CO at fixed PAH intensity for a galaxy correlates with its stellar mass, $M_\star$, and anti-correlates with its specific star formation rate SFR/$M_\star$, in good agreement with results for integrated galaxies using WISE. We provide prescriptions to predict resolved CO emission from PAH emission that also take into account galaxy-dependent normalizations and these represent our best overall predictor.

\item  We also present scaling relations and observe strong correlations linking CO(2-1) and emission in the PAH-dominated F1130W filter as well as continuum-subtracted F335M$_{\rm PAH}$ emission. The F1130W band has coarser resolution but is least affected by starlight contamination out of the three PAH-tracing bands we consider. The fainter F335M$_{\rm PAH}$ emission offers the prospect of tracing CO at thie highest resolution but depends sensitively on stellar continuum subtraction \citep[H. Koziol et al. in preparation;][]{sandstrom2023,bolatto2024}.
\end{enumerate}

\section{Acknowledgments}

This work has been carried out as part of the PHANGS collaboration. This work is based on observations made with the NASA/ESA/CSA JWST. The data were obtained from the Mikulski Archive for Space Telescopes at the Space Telescope Science Institute, which is operated by the Association of Universities for Research in Astronomy, Inc., under NASA contract NAS 5-03127 for JWST. These observations are associated with programs 2107 and 3707. 
  
A.K.L., D.P., S.S., and R.C. gratefully acknowledge support from NSF AST AWD 2205628, JWST-GO-02107.009-A, and JWST-GO-03707.001-A. D.P. is supported by the NSF GRFP. A.K.L. also gratefully acknowledges support by a Humboldt Research Award.

K.S., H.K., and J.S. acknowledge funding support from grants JWST-GO-02107.006-A and JWST-GO-03707.005-A.
JC acknowledges funding from the Belgian Science Policy Office (BELSPO) through the PRODEX project ``JWST/MIRI Science exploitation'' (C4000142239). OE acknowledges funding from the Deutsche Forschungsgemeinschaft (DFG, German Research Foundation) -- project-ID 541068876. 

J.K. is supported by a Kavli Fellowship at the Kavli Institute for Particle Astrophysics and Cosmology (KIPAC).

MB acknowledges support from FONDECYT regular grant 1211000 and by the ANID BASAL project FB210003. This work was supported by the French government through the France 2030 investment plan managed by the National Research Agency (ANR), as part of the Initiative of Excellence of Université Côte d’Azur under reference number ANR-15-IDEX-01.

DC and ZB acknowledge support by the \emph{Deut\-sche For\-schungs\-ge\-mein\-schaft, DFG\/} project number SFB1601-B3.

KK gratefully acknowledges funding from the Deutsche Forschungsgemeinschaft (DFG, German Research Foundation) in the form of an Emmy Noether Research Group (grant number KR4598/2-1, PI Kreckel) and the European Research Council’s starting grant ERC StG-101077573 (“ISM-METALS"). 

RSK acknowledges financial support from ERC via Synergy Grant ``ECOGAL'' (project ID 855130),  from the German Excellence Strategy via the ``STRUCTURES'' Cluster of Excellence (EXC 2181 - 390900948), and from the German Ministry for Economic Affairs and Climate Action in project ``MAINN'' (funding ID 50OO2206). RSK also thanks the 2024/25 Class of Radcliffe Fellows for their company and for highly stimulating discussions.

ER and HH acknowledge support from the Canadian Space Agency, funding reference 23JWGO2A07.

This paper makes use of the following ALMA data, which have been processed as part of the PHANGS--ALMA survey: \\
\noindent ADS/JAO.ALMA\#2012.1.00650.S, \linebreak %
ADS/JAO.ALMA\#2013.1.00803.S, \linebreak %
ADS/JAO.ALMA\#2013.1.01161.S, \linebreak %
ADS/JAO.ALMA\#2015.1.00121.S, \linebreak %
ADS/JAO.ALMA\#2015.1.00782.S, \linebreak %
ADS/JAO.ALMA\#2015.1.00925.S, \linebreak %
ADS/JAO.ALMA\#2015.1.00956.S, \linebreak %
ADS/JAO.ALMA\#2016.1.00386.S, \linebreak %
ADS/JAO.ALMA\#2017.1.00392.S, \linebreak %
ADS/JAO.ALMA\#2017.1.00766.S, \linebreak %
ADS/JAO.ALMA\#2017.1.00886.L, \linebreak %
ADS/JAO.ALMA\#2018.1.01321.S, \linebreak %
ADS/JAO.ALMA\#2018.1.01651.S, \linebreak %
ADS/JAO.ALMA\#2018.A.00062.S, \linebreak %
ADS/JAO.ALMA\#2019.1.01235.S, \linebreak %
ADS/JAO.ALMA\#2019.2.00129.S, \linebreak %
ALMA is a partnership of ESO (representing its member states), NSF (USA), and NINS (Japan), together with NRC (Canada), NSC and ASIAA (Taiwan), and KASI (Republic of Korea), in cooperation with the Republic of Chile. The Joint ALMA Observatory is operated by ESO, AUI/NRAO, and NAOJ. The National Radio Astronomy Observatory is a facility of the National Science Foundation operated under cooperative agreement by Associated Universities, Inc.

Finally, we thank the anonymous referee for their helpful suggestions which improved the quality of this paper.

\facilities{JWST, ALMA, VLT/MUSE}

\software{astropy \citep{astropy-collaboration2013,astropy-collaboration2018, astropy-collaboration2022}}

\appendix

\section{Cycle 2 JWST Imaging of PHANGS Galaxies} 
\label{app:treasury}

During its Cycle 2 campaign, JWST observed 55 nearby galaxies as part of Treasury program GO 3707 ``A JWST Census of the Local Galaxy Population: Anchoring the Physics of the Matter Cycle'' (PI: Leroy, co-PIs: Kreckel, Lee, Rosolowsky, Sandstrom, Schinnerer). The targets were chosen to overlap the PHANGS--ALMA CO~(2-1) survey \citep[][]{leroy2021}, accounting for some updated knowledge about the target selection. PHANGS--ALMA serves as the parent sample for surveys with HST, AstroSat, MeerKAT, VLT-MUSE, and more \citep[][]{emsellem2022,lee2022,hassani2023,eibensteiner2024}. Therefore these observations immediately overlap a rich multiwavelength database that enables a wide variety of science related to PAHs, ISM structure, stellar feedback, star formation, and more.

\startlongtable
\begin{deluxetable*}{lcccccccccccccc}
\tablecaption{PHANGS-JWST Galaxy Sample and Observations}
\label{tab:sample}
\tablehead{
\colhead{Galaxy} & \colhead{$\log M_*$} & \colhead{$\log \mathrm{SFR}/M_*$} & \colhead{Cycle} & \colhead{\rotatebox{90}{F150W}} & \colhead{\rotatebox{90}{F187N}} & \colhead{\rotatebox{90}{F200W}} & \colhead{\rotatebox{90}{F300M}} & \colhead{\rotatebox{90}{F335M}} & \colhead{\rotatebox{90}{F360M}} & \colhead{\rotatebox{90}{F770W}} & \colhead{\rotatebox{90}{F1000W}} & \colhead{\rotatebox{90}{F1130W}} & \colhead{\rotatebox{90}{F2100W}}\\
\colhead{} & \colhead{[$M_\odot$]} & \colhead{[yr$^{-1}$]} & \colhead{} &  &  &  &  &  &  &  &  &  & \\
}
\startdata
IC1954 & $9.67$ & $-10.11$ & $2$ & \checkmark & \checkmark &  & \checkmark & \checkmark &  & \checkmark &  &  & \checkmark \\
IC5273 & $9.72$ & $-9.99$ & $2$ & \checkmark & \checkmark &  & \checkmark & \checkmark &  & \checkmark &  &  & \checkmark \\
IC5332 & $9.67$ & $-10.05$ & $1$ &  &  & \checkmark & \checkmark & \checkmark & \checkmark & \checkmark & \checkmark & \checkmark & \checkmark \\
NGC0628 & $10.34$ & $-10.10$ & $1$ &  &  & \checkmark & \checkmark & \checkmark & \checkmark & \checkmark & \checkmark & \checkmark & \checkmark \\
NGC0685 & $10.06$ & $-10.44$ & $2$ & \checkmark & \checkmark &  & \checkmark & \checkmark &  & \checkmark &  &  & \checkmark \\
NGC1068 & $10.91$ & $-9.27$ & $2$ & \checkmark & \checkmark &  & \checkmark & \checkmark &  & \checkmark &  &  & \checkmark \\
NGC1087 & $9.94$ & $-9.83$ & $1$ &  &  & \checkmark & \checkmark & \checkmark & \checkmark & \checkmark & \checkmark & \checkmark & \checkmark \\
NGC1097 & $10.76$ & $-10.08$ & $2$ & \checkmark & \checkmark &  & \checkmark & \checkmark &  & \checkmark &  &  & \checkmark \\
NGC1300 & $10.62$ & $-10.55$ & $1$ &  &  & \checkmark & \checkmark & \checkmark & \checkmark & \checkmark & \checkmark & \checkmark & \checkmark \\
NGC1317 & $10.62$ & $-10.94$ & $2$ & \checkmark &  & \checkmark & \checkmark & \checkmark &  & \checkmark &  &  & \checkmark \\
NGC1365 & $11.00$ & $-9.76$ & $1$ &  &  & \checkmark & \checkmark & \checkmark & \checkmark & \checkmark & \checkmark & \checkmark & \checkmark \\
NGC1385 & $9.98$ & $-9.66$ & $1$ &  &  & \checkmark & \checkmark & \checkmark & \checkmark & \checkmark & \checkmark & \checkmark & \checkmark \\
NGC1433 & $10.87$ & $-10.82$ & $1$ &  &  & \checkmark & \checkmark & \checkmark & \checkmark & \checkmark & \checkmark & \checkmark & \checkmark \\
NGC1511 & $9.91$ & $-9.55$ & $2$ & \checkmark & \checkmark &  & \checkmark & \checkmark &  & \checkmark &  &  & \checkmark \\
NGC1512 & $10.72$ & $-10.61$ & $1$ &  &  & \checkmark & \checkmark & \checkmark & \checkmark & \checkmark & \checkmark & \checkmark & \checkmark \\
NGC1546 & $10.35$ & $-10.43$ & $2$ & \checkmark & \checkmark &  & \checkmark & \checkmark &  & \checkmark &  &  & \checkmark \\
NGC1559 & $10.36$ & $-9.79$ & $2$ & \checkmark & \checkmark &  & \checkmark & \checkmark &  & \checkmark &  &  & \checkmark \\
NGC1566 & $10.79$ & $-10.13$ & $1$ &  &  & \checkmark & \checkmark & \checkmark & \checkmark & \checkmark & \checkmark & \checkmark & \checkmark \\
NGC1637 & $9.95$ & $-10.14$ & $2$ & \checkmark & \checkmark &  & \checkmark & \checkmark &  & \checkmark &  &  & \checkmark \\
NGC1672 & $10.73$ & $-9.85$ & $1$ &  &  & \checkmark & \checkmark & \checkmark & \checkmark & \checkmark & \checkmark & \checkmark & \checkmark \\
NGC1792 & $10.61$ & $-10.04$ & $2$ & \checkmark & \checkmark &  & \checkmark & \checkmark &  & \checkmark &  &  & \checkmark \\
NGC1808 & $10.29$ & $-9.39$ & $2$ & \checkmark & \checkmark &  & \checkmark & \checkmark &  & \checkmark &  &  & \checkmark \\
NGC1809 & $9.77$ & $-9.01$ & $2$ & \checkmark & \checkmark &  & \checkmark & \checkmark &  & \checkmark &  &  & \checkmark \\
NGC2090 & $10.04$ & $-10.43$ & $2$ & \checkmark & \checkmark &  & \checkmark & \checkmark &  & \checkmark &  &  & \checkmark \\
NGC2283 & $9.89$ & $-10.17$ & $2$ & \checkmark & \checkmark &  & \checkmark & \checkmark &  & \checkmark &  &  & \checkmark \\
NGC2566 & $10.71$ & $-9.77$ & $2$ & \checkmark &  & \checkmark & \checkmark & \checkmark &  & \checkmark &  &  & \checkmark \\
NGC2775 & $11.07$ & $-11.13$ & $2$ & \checkmark & \checkmark &  & \checkmark & \checkmark &  & \checkmark &  &  & \checkmark \\
NGC2835 & $10.00$ & $-9.90$ & $1$ &  &  & \checkmark & \checkmark & \checkmark & \checkmark & \checkmark & \checkmark & \checkmark & \checkmark \\
NGC2903 & $10.63$ & $-10.15$ & $2$ & \checkmark & \checkmark &  & \checkmark & \checkmark &  & \checkmark &  &  & \checkmark \\
NGC2997 & $10.73$ & $-10.09$ & $2$ & \checkmark & \checkmark &  & \checkmark & \checkmark &  & \checkmark &  &  & \checkmark \\
NGC3059 & $10.38$ & $-10.00$ & $2$ & \checkmark & \checkmark &  & \checkmark & \checkmark &  & \checkmark &  &  & \checkmark \\
NGC3137 & $9.88$ & $-10.19$ & $2$ & \checkmark & \checkmark &  & \checkmark & \checkmark &  & \checkmark &  &  & \checkmark \\
NGC3239 & $9.17$ & $-9.58$ & $2$ & \checkmark & \checkmark &  & \checkmark & \checkmark &  & \checkmark &  &  & \checkmark \\
NGC3344 & $10.05$ & $-10.14$ & $2$ & \checkmark & \checkmark &  & \checkmark & \checkmark &  & \checkmark &  &  & \checkmark \\
NGC3351 & $10.37$ & $-10.25$ & $1$ &  &  & \checkmark & \checkmark & \checkmark & \checkmark & \checkmark & \checkmark & \checkmark & \checkmark \\
NGC3368 & $10.74$ & $-10.88$ & $2$ & \checkmark & \checkmark &  & \checkmark & \checkmark &  & \checkmark &  &  & \checkmark \\
NGC3507 & $10.40$ & $-10.40$ & $2$ & \checkmark & \checkmark &  & \checkmark & \checkmark &  & \checkmark &  &  & \checkmark \\
NGC3511 & $10.03$ & $-10.12$ & $2$ & \checkmark & \checkmark &  & \checkmark & \checkmark &  & \checkmark &  &  & \checkmark \\
NGC3521 & $11.02$ & $-10.45$ & $2$ & \checkmark & \checkmark &  & \checkmark & \checkmark &  & \checkmark &  &  & \checkmark \\
NGC3596 & $9.66$ & $-10.18$ & $2$ & \checkmark & \checkmark &  & \checkmark & \checkmark &  & \checkmark &  &  & \checkmark \\
NGC3621 & $10.06$ & $-10.06$ & $2$ & \checkmark & \checkmark &  & \checkmark & \checkmark &  & \checkmark &  &  & \checkmark \\
NGC3626 & $10.46$ & $-11.13$ & $2$ & \checkmark &  & \checkmark & \checkmark & \checkmark &  & \checkmark &  &  & \checkmark \\
NGC3627 & $10.84$ & $-10.25$ & $1$ &  &  & \checkmark & \checkmark & \checkmark & \checkmark & \checkmark & \checkmark & \checkmark & \checkmark \\
NGC4254 & $10.42$ & $-9.93$ & $1$ &  &  & \checkmark & \checkmark & \checkmark & \checkmark & \checkmark & \checkmark & \checkmark & \checkmark \\
NGC4298 & $10.02$ & $-10.36$ & $2$ & \checkmark & \checkmark &  & \checkmark & \checkmark &  & \checkmark &  &  & \checkmark \\
NGC4303 & $10.51$ & $-9.78$ & $1$ &  &  & \checkmark & \checkmark & \checkmark & \checkmark & \checkmark & \checkmark & \checkmark & \checkmark \\
NGC4321 & $10.75$ & $-10.20$ & $1$ &  &  & \checkmark & \checkmark & \checkmark & \checkmark & \checkmark & \checkmark & \checkmark & \checkmark \\
NGC4424 & $9.91$ & $-10.43$ & $2$ & \checkmark & \checkmark &  & \checkmark & \checkmark &  & \checkmark &  &  & \checkmark \\
NGC4457 & $10.42$ & $-10.93$ & $2$ & \checkmark & \checkmark &  & \checkmark & \checkmark &  & \checkmark &  &  & \checkmark \\
NGC4496A & $9.53$ & $-9.74$ & $2$ & \checkmark &  & \checkmark & \checkmark & \checkmark &  & \checkmark &  &  & \checkmark \\
NGC4535 & $10.54$ & $-10.20$ & $1$ &  &  & \checkmark & \checkmark & \checkmark & \checkmark & \checkmark & \checkmark & \checkmark & \checkmark \\
NGC4536 & $10.40$ & $-9.86$ & $2$ & \checkmark &  & \checkmark & \checkmark & \checkmark &  & \checkmark &  &  & \checkmark \\
NGC4540 & $9.79$ & $-10.56$ & $2$ & \checkmark & \checkmark &  & \checkmark & \checkmark &  & \checkmark &  &  & \checkmark \\
NGC4548 & $10.69$ & $-10.97$ & $2$ & \checkmark & \checkmark &  & \checkmark & \checkmark &  & \checkmark &  &  & \checkmark \\
NGC4569 & $10.81$ & $-10.68$ & $2$ & \checkmark & \checkmark &  & \checkmark & \checkmark &  & \checkmark &  &  & \checkmark \\
NGC4571 & $10.09$ & $-10.63$ & $2$ & \checkmark & \checkmark &  & \checkmark & \checkmark &  & \checkmark &  &  & \checkmark \\
NGC4579 & $11.15$ & $-10.81$ & $2$ & \checkmark &  & \checkmark & \checkmark & \checkmark &  & \checkmark &  &  & \checkmark \\
NGC4654 & $10.57$ & $-9.99$ & $2$ & \checkmark & \checkmark &  & \checkmark & \checkmark &  & \checkmark &  &  & \checkmark \\
NGC4689 & $10.22$ & $-10.61$ & $2$ & \checkmark &  & \checkmark & \checkmark & \checkmark &  & \checkmark &  &  & \checkmark \\
NGC4694 & $9.86$ & $-10.66$ & $2$ & \checkmark & \checkmark &  & \checkmark & \checkmark &  & \checkmark &  &  & \checkmark \\
NGC4731 & $9.48$ & $-9.70$ & $2$ & \checkmark &  & \checkmark & \checkmark & \checkmark &  & \checkmark &  &  & \checkmark \\
NGC4781 & $9.64$ & $-9.96$ & $2$ & \checkmark & \checkmark &  & \checkmark & \checkmark &  & \checkmark &  &  & \checkmark \\
NGC4826 & $10.24$ & $-10.93$ & $2$ & \checkmark & \checkmark &  & \checkmark & \checkmark &  & \checkmark &  &  & \checkmark \\
NGC4941 & $10.17$ & $-10.53$ & $2$ & \checkmark & \checkmark &  & \checkmark & \checkmark &  & \checkmark &  &  & \checkmark \\
NGC4951 & $9.79$ & $-10.24$ & $2$ & \checkmark & \checkmark &  & \checkmark & \checkmark &  & \checkmark &  &  & \checkmark \\
NGC5042 & $9.90$ & $-10.12$ & $2$ & \checkmark & \checkmark &  & \checkmark & \checkmark &  & \checkmark &  &  & \checkmark \\
NGC5068 & $9.41$ & $-9.97$ & $1$ &  &  & \checkmark & \checkmark & \checkmark & \checkmark & \checkmark & \checkmark & \checkmark & \checkmark \\
NGC5134 & $10.41$ & $-10.75$ & $2$ & \checkmark &  & \checkmark & \checkmark & \checkmark &  & \checkmark &  &  & \checkmark \\
NGC5248 & $10.41$ & $-10.05$ & $2$ & \checkmark & \checkmark &  & \checkmark & \checkmark &  & \checkmark &  &  & \checkmark \\
NGC5643 & $10.34$ & $-9.92$ & $2$ & \checkmark & \checkmark &  & \checkmark & \checkmark &  & \checkmark &  &  & \checkmark \\
NGC6300 & $10.47$ & $-10.19$ & $2$ & \checkmark & \checkmark &  & \checkmark & \checkmark &  & \checkmark &  &  & \checkmark \\
NGC7456 & $9.64$ & $-10.08$ & $2$ & \checkmark & \checkmark &  & \checkmark & \checkmark &  & \checkmark &  &  & \checkmark \\
NGC7496 & $10.00$ & $-9.65$ & $1$ &  &  & \checkmark & \checkmark & \checkmark & \checkmark & \checkmark & \checkmark & \checkmark & \checkmark \\
\hline
Total & --- & --- & --- & $54$ & $45$ & $28$ & $73$ & $73$ & $19$ & $73$ & $19$ & $19$ & $73$ \\
\enddata
\tablecomments{Columns: $\log M_*$ --- global stellar mass; $\log \mathrm{SFR}/M_*$ --- global specific star formation rate; Cycle --- JWST observing cycle when the galaxy was observed (Cycle 1 corresponds to program \# 2107; Cycle 2 corresponds to program \# 3707); Remaining columns show which JWST bands have observations (\checkmark) for each galaxy. The bottom row shows the total number of galaxies observed in each band.}
\end{deluxetable*}

\begin{figure*}
\begin{center}
\includegraphics[width=0.5\textwidth]{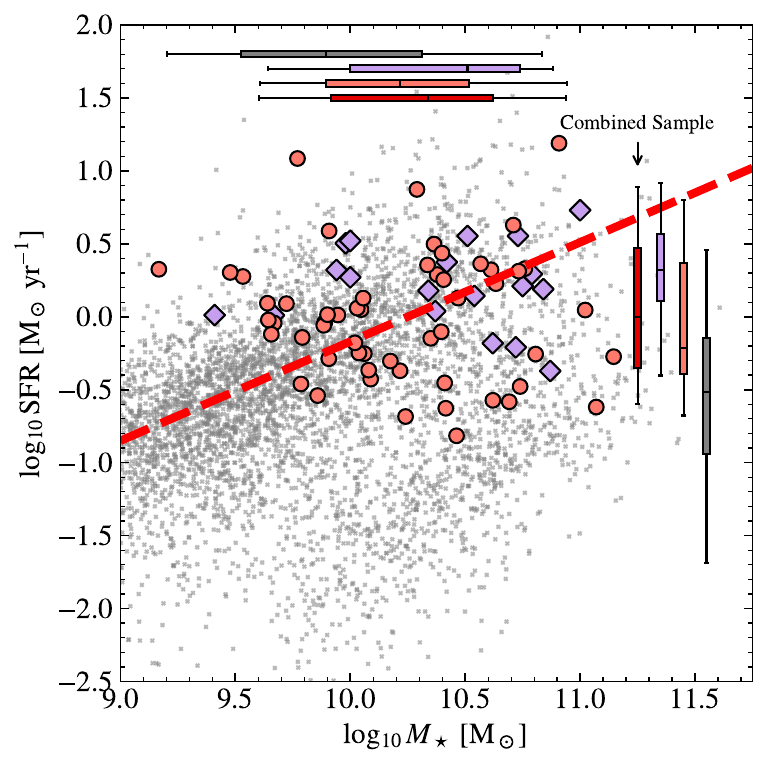}\includegraphics[width=0.5\textwidth]{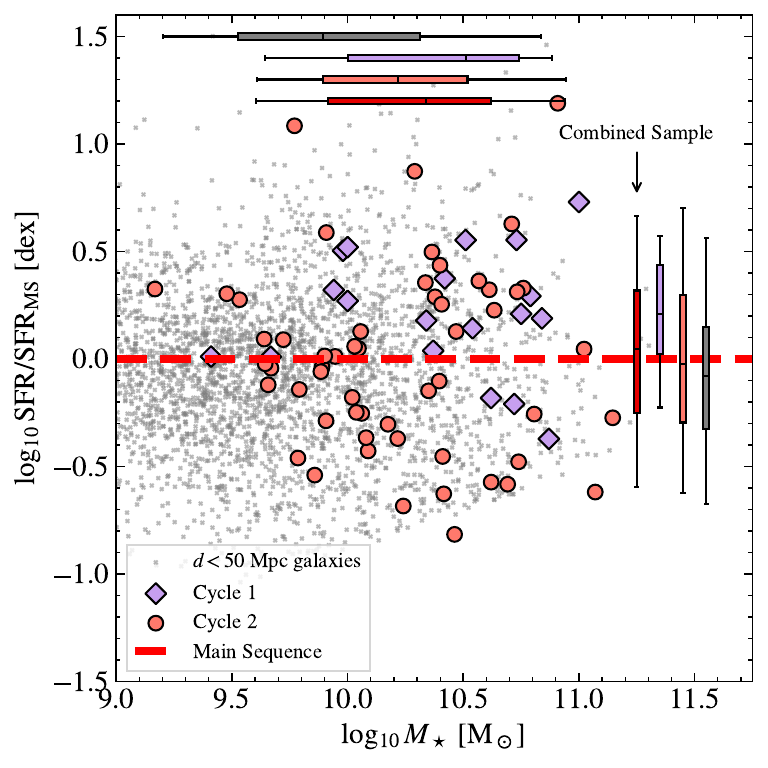}
\caption{\textbf{PHANGS-JWST samples relative to the star-forming main sequence.} The reference line in the left panel is derived from Eq. 19 of \citet{leroy2019}, with a slope of $0.68$ and a y-intercept of $-6.97~\mathrm{M_\odot~yr^{-1}}$. The box plots show the median, 16th and 84th percentiles, while the whiskers show the 5th and 95th percentiles. The red box and whiskers show that Cycle 1 + Cycle 2 PHANGS-JWST galaxies cover about 1 dex in stellar mass and approximately $\pm 0.5$ dex about the main sequence. The Cycle 1 targets have higher masses and star formation rates than the rest of the sample. }
\label{fig:sample}
\end{center}
\end{figure*}

Table \ref{tab:sample} lists the targets of both the Cycle 1 and Cycle 2 PHANGS Treasury, along with the list of bands observed and the stellar mass and SFR for each target \citep[mostly from][]{leroy2019,leroy2021}. Figure \ref{fig:sample} shows both samples in SFR-M$_\star$ parameter space. As of this writing, all but two of the Cycle 2 galaxies (NGC 5248 and NGC 5530) have been fully observed. Those two require partial repeated observations. The observations for NGC~5248 and NGC~5530 are expected to be completed by mid-July 2025 and late-June 2025, respectively.

The Cycle 2 survey complements the one described in \citet{lee2023} and \citet{williams2024}. This new survey targets 55 galaxies, expanding on the 19 in the original sample. Compared to the original sample, these new observations expand to cover galaxies with lower mass and lower specific star formation rate. Taken together, the two surveys sample the main sequence of star-forming galaxies over the range $\log M_\star / M_\odot \approx 9.75{-}10.75$, i.e., within about $1$~dex of the knee in the galaxy mass function and covering the main region of SFR-$M_\star$ parameter space (Figure~\ref{fig:sample}) where stars form at $z=0$ \citep[see][]{leroy2021}. In order to cover this large sample, the current survey focused on two PAH-sensitive filters (F335M and F770W) and did not use the F1130W filter employed by \citet{lee2023}. The Cycle 2 survey also did not observe the F360M filter that was included by \citet{lee2023}, and we shifted the short wavelength starlight-sensitive filter from F200W to F150W in order to avoid wavelength overlap with the F187N filter. We added the F187N filter (which was not observed by \citet{lee2023}) wherever it captures the Paschen-$\alpha$ line. This yields Paschen-$\alpha$ imaging, which should open a powerful new scientific space (high resolution imaging of recent star formation). 

The observations consisted of MIRI mosaics using the F770W and F2100W filters, with $\approx 88$s exposure time per point at F770W and $\approx 344$s at F2100W. The coverage of each mosaic was designed to match the extent of the galaxy in the $\approx 12\mu$m emission captured by WISE band 3 \citep[][]{wright2010}. Roll angle constraints, row offsets, and other aspects of the observation were designed to meet this goal. As in \citet{lee2023} and \citet{williams2024}, the  MIRI observations were paired in a non-interruptible sequence with single-field off-galaxy observations designed to measure the local background. 

In parallel with the background observations, we observed the center of the galaxy with NIRCam module B using the F150W (integration time 215s), F187N (covering the Paschen $\alpha$ line, 386s), F300M (215s), and F335M (covering the $3.3\mu$m PAH feature, 386s) filters. In a few cases where the redshift of the galaxy was too large for the F187N filter to capture the Paschen $\alpha$ line, we observed using the F200W filter instead. The $2'.2 \times 2'.2$ NIRCam field covers a large portion of each galaxy ($\approx 6 \times 6$~kpc at $10$~Mpc and $\approx 12 \times 12$~kpc at $20$~Mpc). The offset between NIRCam and MIRI was enough to allow the MIRI observations to reach a sky position off the galaxy. 

As discussed in \S \ref{sec:data}, we processed the data using \texttt{pjpipe}, the modified version of the STScI pipeline described in \citet{williams2024} and \citet{lee2023}. Refinements to the procedures involved in \citet{williams2024} are ongoing. As of this writing, key difference compared to that paper include: (1) for MIRI, we have masked out the Lyot coronagraph as we find this to improve the flux-level match between overlapping tiles, (2) improvements in the ``anchoring'' \citep{leroy2023} of the background level to reference data (H. Koziol et al. in preparation, D. Pathak et al. in preparation), (3) improved cross-registration between filters (especially relevant to the $3.3\mu$m and Paschen-$\alpha$ analyses; H. Koziol et al., in preparation, T. Weinbeck et al., in preparation), and (4) improved techniques to remove $1/f$-related artifacts from the F187N images (T. Weinbeck et al. in preparation).

\section{Fits for individual galaxies}\label{appendix:allgals}

Table~\ref{tab:galtab} shows the best-fit parameters of the CO(2-1) vs. F335M$_\mathrm{PAH}$, F770W$_\mathrm{PAH}$, and F1130W relationships for each galaxy. We also report the stellar mass and SFR/$M_\star$, adopted from \citet{leroy2019}, the brightness of band $X$ in that galaxy at $I_{\rm \nu}^X = 1$~MJy~sr$^{-1}$, the number of sight lines analyzed, and Spearman's rank correlation coefficient relating the PAH and CO emission for each sightline.

\startlongtable
\begin{deluxetable*}{lrrrrrrrrrr}
\label{tab:galtab}
\tablecolumns{10}
\tablecaption{Fits to all pixels outside galaxy centers for each galaxy.}
\tablehead{
\colhead{Name} & \colhead{Cycle} & \colhead{$\log_{10}M_\star$} & \colhead{$\log_{10}\mathrm{SFR}/M_\star$} & \colhead{$m$} & \colhead{$b$} & \colhead{$x_0$} & \colhead{$\log_{10}C_\mathrm{band}$} & \colhead{$N_\mathrm{pix}$} & \colhead{$\sigma_\mathrm{pix}$} & \colhead{$r$}}
\startdata
\cutinhead{$I_\mathrm{CO(2-1)}$ vs F770W$_\mathrm{PAH}$}
IC1954 & 2 & $9.67$ & $-10.11$ & $1.050 \pm 0.017$ & $0.199 \pm 0.010$ & $0.366$ & $-0.185$ & $16101$ & $0.09$ & $0.76$ \\
IC5273 & 2 & $9.72$ & $-9.99$ & $1.133 \pm 0.046$ & $-0.069 \pm 0.023$ & $0.226$ & $-0.325$ & $13117$ & $0.13$ & $0.66$ \\
IC5332 & 2 & $9.67$ & $-10.05$ & $0.438 \pm 0.010$ & $0.261 \pm 0.005$ & $-0.079$ & $0.296$ & $16281$ & $0.20$ & $0.22$ \\
NGC0628 & 1 & $10.34$ & $-10.10$ & $0.689 \pm 0.010$ & $0.377 \pm 0.008$ & $0.179$ & $0.254$ & $122222$ & $0.16$ & $0.52$ \\
NGC0685 & 2 & $10.06$ & $-10.44$ & $0.986 \pm 0.020$ & $-0.083 \pm 0.012$ & $-0.027$ & $-0.056$ & $13882$ & $0.16$ & $0.53$ \\
NGC1087 & 1 & $9.94$ & $-9.83$ & $1.046 \pm 0.024$ & $0.290 \pm 0.021$ & $0.466$ & $-0.198$ & $26830$ & $0.15$ & $0.73$ \\
NGC1097 & 2 & $10.76$ & $-10.08$ & $1.238 \pm 0.033$ & $0.307 \pm 0.016$ & $0.155$ & $0.115$ & $24368$ & $0.22$ & $0.57$ \\
NGC1300 & 1 & $10.62$ & $-10.55$ & $0.822 \pm 0.025$ & $0.189 \pm 0.017$ & $0.042$ & $0.155$ & $36359$ & $0.20$ & $0.47$ \\
NGC1317 & 2 & $10.62$ & $-10.94$ & $0.861 \pm 0.044$ & $0.816 \pm 0.023$ & $0.738$ & $0.181$ & $4659$ & $0.13$ & $0.80$ \\
NGC1365 & 1 & $11.00$ & $-9.76$ & $0.983 \pm 0.051$ & $0.690 \pm 0.053$ & $0.292$ & $0.403$ & $40810$ & $0.25$ & $0.46$ \\
NGC1385 & 1 & $9.98$ & $-9.66$ & $0.885 \pm 0.019$ & $0.377 \pm 0.015$ & $0.560$ & $-0.118$ & $34515$ & $0.17$ & $0.72$ \\
NGC1433 & 1 & $10.87$ & $-10.82$ & $1.176 \pm 0.070$ & $0.119 \pm 0.048$ & $-0.069$ & $0.200$ & $19809$ & $0.24$ & $0.50$ \\
NGC1511 & 2 & $9.91$ & $-9.55$ & $0.904 \pm 0.018$ & $0.623 \pm 0.013$ & $0.811$ & $-0.111$ & $12884$ & $0.11$ & $0.78$ \\
NGC1512 & 1 & $10.72$ & $-10.61$ & $0.833 \pm 0.027$ & $0.172 \pm 0.014$ & $0.009$ & $0.164$ & $14121$ & $0.16$ & $0.41$ \\
NGC1546 & 2 & $10.35$ & $-10.43$ & $0.916 \pm 0.023$ & $0.961 \pm 0.013$ & $0.848$ & $0.184$ & $11723$ & $0.05$ & $0.91$ \\
NGC1559 & 2 & $10.36$ & $-9.79$ & $0.943 \pm 0.025$ & $0.537 \pm 0.017$ & $0.763$ & $-0.182$ & $46456$ & $0.16$ & $0.70$ \\
NGC1566 & 1 & $10.79$ & $-10.13$ & $0.905 \pm 0.015$ & $0.366 \pm 0.014$ & $0.335$ & $0.063$ & $87712$ & $0.22$ & $0.68$ \\
NGC1637 & 2 & $9.95$ & $-10.14$ & $1.028 \pm 0.029$ & $0.124 \pm 0.018$ & $0.121$ & $-0.001$ & $39715$ & $0.15$ & $0.66$ \\
NGC1672 & 1 & $10.73$ & $-9.85$ & $1.066 \pm 0.030$ & $0.364 \pm 0.021$ & $0.371$ & $-0.031$ & $23689$ & $0.20$ & $0.66$ \\
NGC1792 & 2 & $10.61$ & $-10.04$ & $1.097 \pm 0.025$ & $0.775 \pm 0.015$ & $0.896$ & $-0.209$ & $24488$ & $0.12$ & $0.84$ \\
NGC1808 & 2 & $10.29$ & $-9.39$ & $1.103 \pm 0.081$ & $0.983 \pm 0.069$ & $0.879$ & $0.014$ & $10521$ & $0.24$ & $0.78$ \\
NGC1809 & 2 & $9.77$ & $-9.01$ & $1.134 \pm 0.038$ & $0.085 \pm 0.016$ & $0.223$ & $-0.169$ & $7076$ & $0.17$ & $0.53$ \\
NGC2090 & 2 & $10.04$ & $-10.43$ & $0.913 \pm 0.062$ & $0.493 \pm 0.024$ & $0.404$ & $0.124$ & $15649$ & $0.11$ & $0.63$ \\
NGC2283 & 2 & $9.89$ & $-10.17$ & $0.927 \pm 0.047$ & $0.080 \pm 0.027$ & $0.261$ & $-0.162$ & $26500$ & $0.19$ & $0.52$ \\
NGC2566 & 2 & $10.71$ & $-9.77$ & $0.987 \pm 0.018$ & $0.307 \pm 0.011$ & $0.230$ & $0.080$ & $35872$ & $0.24$ & $0.57$ \\
NGC2775 & 2 & $11.07$ & $-11.13$ & $0.856 \pm 0.023$ & $0.355 \pm 0.009$ & $0.138$ & $0.237$ & $44187$ & $0.13$ & $0.47$ \\
NGC2835 & 1 & $10.00$ & $-9.90$ & $0.790 \pm 0.021$ & $0.172 \pm 0.015$ & $0.149$ & $0.054$ & $36639$ & $0.20$ & $0.42$ \\
NGC2903 & 2 & $10.63$ & $-10.15$ & $0.944 \pm 0.026$ & $0.623 \pm 0.019$ & $0.674$ & $-0.013$ & $64047$ & $0.14$ & $0.73$ \\
NGC2997 & 2 & $10.73$ & $-10.09$ & $1.050 \pm 0.012$ & $0.349 \pm 0.008$ & $0.267$ & $0.069$ & $44985$ & $0.15$ & $0.70$ \\
NGC3059 & 2 & $10.38$ & $-10.00$ & $0.939 \pm 0.017$ & $0.293 \pm 0.018$ & $0.355$ & $-0.040$ & $61109$ & $0.18$ & $0.64$ \\
NGC3137 & 2 & $9.88$ & $-10.19$ & $0.956 \pm 0.067$ & $0.354 \pm 0.020$ & $0.348$ & $0.021$ & $4956$ & $0.08$ & $0.53$ \\
NGC3239 & 2 & $9.17$ & $-9.58$ & $0.573 \pm 0.048$ & $0.073 \pm 0.016$ & $0.222$ & $-0.055$ & $1640$ & $0.17$ & $0.30$ \\
NGC3351 & 1 & $10.37$ & $-10.25$ & $0.650 \pm 0.016$ & $0.320 \pm 0.008$ & $0.051$ & $0.287$ & $29148$ & $0.17$ & $0.33$ \\
NGC3507 & 2 & $10.40$ & $-10.40$ & $0.907 \pm 0.020$ & $0.127 \pm 0.011$ & $0.026$ & $0.103$ & $32339$ & $0.17$ & $0.49$ \\
NGC3511 & 2 & $10.03$ & $-10.12$ & $1.185 \pm 0.025$ & $0.514 \pm 0.011$ & $0.645$ & $-0.250$ & $11678$ & $0.08$ & $0.76$ \\
NGC3521 & 2 & $11.02$ & $-10.45$ & $1.155 \pm 0.031$ & $0.869 \pm 0.017$ & $1.001$ & $-0.287$ & $69658$ & $0.10$ & $0.85$ \\
NGC3596 & 2 & $9.66$ & $-10.18$ & $0.814 \pm 0.019$ & $0.349 \pm 0.012$ & $0.236$ & $0.157$ & $36060$ & $0.17$ & $0.57$ \\
NGC3621 & 2 & $10.06$ & $-10.06$ & $1.161 \pm 0.021$ & $0.561 \pm 0.012$ & $0.761$ & $-0.322$ & $38133$ & $0.13$ & $0.75$ \\
NGC3626 & 2 & $10.46$ & $-11.13$ & $0.814 \pm 0.036$ & $0.572 \pm 0.014$ & $0.509$ & $0.158$ & $4036$ & $0.14$ & $0.62$ \\
NGC3627 & 1 & $10.84$ & $-10.25$ & $0.993 \pm 0.013$ & $0.743 \pm 0.012$ & $0.736$ & $0.012$ & $67198$ & $0.18$ & $0.72$ \\
NGC4254 & 1 & $10.42$ & $-9.93$ & $1.016 \pm 0.019$ & $0.562 \pm 0.013$ & $0.570$ & $-0.017$ & $65123$ & $0.14$ & $0.78$ \\
NGC4298 & 2 & $10.02$ & $-10.36$ & $0.967 \pm 0.046$ & $0.518 \pm 0.021$ & $0.377$ & $0.153$ & $20937$ & $0.07$ & $0.75$ \\
NGC4303 & 1 & $10.51$ & $-9.78$ & $0.844 \pm 0.019$ & $0.644 \pm 0.013$ & $0.660$ & $0.086$ & $29397$ & $0.18$ & $0.67$ \\
NGC4321 & 1 & $10.75$ & $-10.20$ & $0.935 \pm 0.007$ & $0.479 \pm 0.005$ & $0.282$ & $0.215$ & $37681$ & $0.16$ & $0.63$ \\
NGC4424 & 2 & $9.91$ & $-10.43$ & $0.816 \pm 0.064$ & $0.406 \pm 0.048$ & $0.480$ & $0.014$ & $4800$ & $0.23$ & $0.57$ \\
NGC4457 & 2 & $10.42$ & $-10.93$ & $0.646 \pm 0.051$ & $0.537 \pm 0.030$ & $0.124$ & $0.457$ & $14785$ & $0.21$ & $0.44$ \\
NGC4496A & 2 & $9.53$ & $-9.74$ & $0.759 \pm 0.024$ & $0.170 \pm 0.014$ & $0.161$ & $0.048$ & $12392$ & $0.18$ & $0.41$ \\
NGC4535 & 1 & $10.54$ & $-10.20$ & $0.971 \pm 0.038$ & $0.338 \pm 0.029$ & $0.158$ & $0.184$ & $31294$ & $0.17$ & $0.60$ \\
NGC4536 & 2 & $10.40$ & $-9.86$ & $1.110 \pm 0.023$ & $0.339 \pm 0.020$ & $0.350$ & $-0.049$ & $17043$ & $0.15$ & $0.74$ \\
NGC4540 & 2 & $9.79$ & $-10.56$ & $0.788 \pm 0.046$ & $0.392 \pm 0.021$ & $0.276$ & $0.174$ & $10939$ & $0.16$ & $0.48$ \\
NGC4548 & 2 & $10.69$ & $-10.97$ & $1.002 \pm 0.020$ & $0.210 \pm 0.010$ & $0.014$ & $0.196$ & $8294$ & $0.16$ & $0.54$ \\
NGC4569 & 2 & $10.81$ & $-10.68$ & $0.841 \pm 0.018$ & $0.837 \pm 0.011$ & $0.598$ & $0.334$ & $14984$ & $0.11$ & $0.63$ \\
NGC4571 & 2 & $10.09$ & $-10.63$ & $0.548 \pm 0.019$ & $0.280 \pm 0.009$ & $-0.029$ & $0.296$ & $26226$ & $0.14$ & $0.28$ \\
NGC4579 & 2 & $11.15$ & $-10.81$ & $0.833 \pm 0.039$ & $0.356 \pm 0.022$ & $0.106$ & $0.268$ & $25120$ & $0.15$ & $0.54$ \\
NGC4654 & 2 & $10.57$ & $-9.99$ & $1.051 \pm 0.027$ & $0.474 \pm 0.020$ & $0.512$ & $-0.064$ & $27373$ & $0.13$ & $0.78$ \\
NGC4689 & 2 & $10.22$ & $-10.61$ & $0.861 \pm 0.021$ & $0.356 \pm 0.013$ & $0.133$ & $0.241$ & $45635$ & $0.14$ & $0.59$ \\
NGC4694 & 2 & $9.86$ & $-10.66$ & $0.821 \pm 0.059$ & $0.294 \pm 0.044$ & $0.360$ & $-0.002$ & $2720$ & $0.21$ & $0.54$ \\
NGC4731 & 2 & $9.48$ & $-9.70$ & $0.937 \pm 0.028$ & $0.002 \pm 0.014$ & $0.279$ & $-0.259$ & $4894$ & $0.18$ & $0.51$ \\
NGC4781 & 2 & $9.64$ & $-9.96$ & $1.064 \pm 0.013$ & $0.265 \pm 0.008$ & $0.491$ & $-0.257$ & $29714$ & $0.12$ & $0.75$ \\
NGC4826 & 2 & $10.24$ & $-10.93$ & $1.015 \pm 0.044$ & $1.084 \pm 0.027$ & $1.013$ & $0.056$ & $22873$ & $0.11$ & $0.83$ \\
NGC4941 & 2 & $10.17$ & $-10.53$ & $0.795 \pm 0.024$ & $0.261 \pm 0.006$ & $0.034$ & $0.234$ & $5447$ & $0.07$ & $0.37$ \\
NGC4951 & 2 & $9.79$ & $-10.24$ & $1.092 \pm 0.039$ & $0.372 \pm 0.014$ & $0.480$ & $-0.153$ & $6724$ & $0.16$ & $0.55$ \\
NGC5042 & 2 & $9.90$ & $-10.12$ & $0.891 \pm 0.018$ & $0.094 \pm 0.008$ & $0.109$ & $-0.003$ & $15685$ & $0.19$ & $0.40$ \\
NGC5068 & 1 & $9.41$ & $-9.97$ & $0.690 \pm 0.021$ & $0.282 \pm 0.016$ & $0.182$ & $0.156$ & $67372$ & $0.23$ & $0.39$ \\
NGC5134 & 2 & $10.41$ & $-10.75$ & $0.737 \pm 0.025$ & $0.196 \pm 0.013$ & $0.117$ & $0.110$ & $19250$ & $0.20$ & $0.44$ \\
NGC5248 & 2 & $10.41$ & $-10.05$ & $0.962 \pm 0.020$ & $0.503 \pm 0.014$ & $0.433$ & $0.087$ & $51724$ & $0.14$ & $0.71$ \\
NGC5643 & 2 & $10.34$ & $-9.92$ & $0.853 \pm 0.015$ & $0.385 \pm 0.010$ & $0.363$ & $0.076$ & $76446$ & $0.19$ & $0.61$ \\
NGC6300 & 2 & $10.47$ & $-10.19$ & $0.905 \pm 0.017$ & $0.481 \pm 0.012$ & $0.383$ & $0.134$ & $84152$ & $0.19$ & $0.58$ \\
NGC7456 & 2 & $9.64$ & $-10.08$ & $0.638 \pm 0.114$ & $0.138 \pm 0.016$ & $0.200$ & $0.011$ & $907$ & $0.06$ & $0.26$ \\
NGC7496 & 1 & $10.00$ & $-9.65$ & $1.032 \pm 0.013$ & $0.062 \pm 0.009$ & $0.137$ & $-0.080$ & $15593$ & $0.15$ & $0.68$ \\
\cutinhead{$I_\mathrm{CO(2-1)}$ vs F335M$_\mathrm{PAH}$}
IC5332 & 2 & $9.67$ & $-10.05$ & $0.339 \pm 0.046$ & $0.416 \pm 0.013$ & $-0.813$ & $0.691$ & $2242$ & $0.18$ & $0.14$ \\
NGC0628 & 1 & $10.34$ & $-10.10$ & $0.630 \pm 0.017$ & $0.623 \pm 0.012$ & $-0.780$ & $1.115$ & $27463$ & $0.11$ & $0.42$ \\
NGC1087 & 1 & $9.94$ & $-9.83$ & $1.211 \pm 0.018$ & $0.401 \pm 0.013$ & $-0.545$ & $1.062$ & $19279$ & $0.12$ & $0.73$ \\
NGC1300 & 1 & $10.62$ & $-10.55$ & $0.862 \pm 0.041$ & $0.589 \pm 0.016$ & $-0.703$ & $1.196$ & $5452$ & $0.12$ & $0.50$ \\
NGC1365 & 1 & $11.00$ & $-9.76$ & $1.295 \pm 0.114$ & $0.842 \pm 0.080$ & $-0.526$ & $1.524$ & $5781$ & $0.33$ & $0.37$ \\
NGC1385 & 1 & $9.98$ & $-9.66$ & $0.924 \pm 0.015$ & $0.632 \pm 0.010$ & $-0.387$ & $0.989$ & $19238$ & $0.11$ & $0.72$ \\
NGC1433 & 1 & $10.87$ & $-10.82$ & $1.604 \pm 0.280$ & $0.776 \pm 0.062$ & $-0.746$ & $1.972$ & $2224$ & $0.32$ & $0.37$ \\
NGC1512 & 1 & $10.72$ & $-10.61$ & $0.714 \pm 0.061$ & $0.538 \pm 0.015$ & $-0.734$ & $1.063$ & $1468$ & $0.08$ & $0.40$ \\
NGC1566 & 1 & $10.79$ & $-10.13$ & $0.793 \pm 0.013$ & $0.770 \pm 0.009$ & $-0.569$ & $1.221$ & $28985$ & $0.21$ & $0.52$ \\
NGC1672 & 1 & $10.73$ & $-9.85$ & $1.126 \pm 0.055$ & $0.657 \pm 0.028$ & $-0.576$ & $1.305$ & $8502$ & $0.20$ & $0.52$ \\
NGC2835 & 1 & $10.00$ & $-9.90$ & $0.803 \pm 0.041$ & $0.483 \pm 0.018$ & $-0.663$ & $1.015$ & $6072$ & $0.15$ & $0.42$ \\
NGC3351 & 1 & $10.37$ & $-10.25$ & $0.722 \pm 0.109$ & $0.615 \pm 0.025$ & $-0.726$ & $1.139$ & $963$ & $0.23$ & $0.21$ \\
NGC3627 & 1 & $10.84$ & $-10.25$ & $1.087 \pm 0.022$ & $0.785 \pm 0.018$ & $-0.428$ & $1.250$ & $54452$ & $0.19$ & $0.64$ \\
NGC4254 & 1 & $10.42$ & $-9.93$ & $0.905 \pm 0.011$ & $0.893 \pm 0.006$ & $-0.500$ & $1.345$ & $32222$ & $0.10$ & $0.66$ \\
NGC4303 & 1 & $10.51$ & $-9.78$ & $0.592 \pm 0.012$ & $0.999 \pm 0.007$ & $-0.466$ & $1.275$ & $9011$ & $0.10$ & $0.54$ \\
NGC4321 & 1 & $10.75$ & $-10.20$ & $0.887 \pm 0.017$ & $0.818 \pm 0.009$ & $-0.667$ & $1.409$ & $8440$ & $0.10$ & $0.59$ \\
NGC4535 & 1 & $10.54$ & $-10.20$ & $0.841 \pm 0.025$ & $0.847 \pm 0.010$ & $-0.611$ & $1.362$ & $4411$ & $0.13$ & $0.50$ \\
NGC5068 & 1 & $9.41$ & $-9.97$ & $0.704 \pm 0.030$ & $0.460 \pm 0.017$ & $-0.679$ & $0.938$ & $26386$ & $0.21$ & $0.36$ \\
NGC7496 & 1 & $10.00$ & $-9.65$ & $0.930 \pm 0.048$ & $0.490 \pm 0.021$ & $-0.695$ & $1.136$ & $3755$ & $0.13$ & $0.54$ \\
\cutinhead{$I_\mathrm{CO(2-1)}$ vs F1130W}
IC5332 & 2 & $9.67$ & $-10.05$ & $0.495 \pm 0.022$ & $0.251 \pm 0.012$ & $0.107$ & $0.199$ & $16263$ & $0.20$ & $0.23$ \\
NGC0628 & 1 & $10.34$ & $-10.10$ & $0.538 \pm 0.054$ & $0.513 \pm 0.044$ & $0.390$ & $0.304$ & $122196$ & $0.20$ & $0.53$ \\
NGC1087 & 1 & $9.94$ & $-9.83$ & $1.034 \pm 0.038$ & $0.312 \pm 0.032$ & $0.637$ & $-0.347$ & $26790$ & $0.16$ & $0.73$ \\
NGC1300 & 1 & $10.62$ & $-10.55$ & $0.617 \pm 0.078$ & $0.339 \pm 0.058$ & $0.228$ & $0.199$ & $36319$ & $0.23$ & $0.51$ \\
NGC1365 & 1 & $11.00$ & $-9.76$ & $1.008 \pm 0.060$ & $0.795 \pm 0.051$ & $0.551$ & $0.240$ & $40793$ & $0.24$ & $0.50$ \\
NGC1385 & 1 & $9.98$ & $-9.66$ & $0.705 \pm 0.055$ & $0.512 \pm 0.050$ & $0.725$ & $0.000$ & $34494$ & $0.21$ & $0.72$ \\
NGC1433 & 1 & $10.87$ & $-10.82$ & $1.060 \pm 0.059$ & $0.256 \pm 0.038$ & $0.200$ & $0.043$ & $19802$ & $0.22$ & $0.57$ \\
NGC1512 & 1 & $10.72$ & $-10.61$ & $0.874 \pm 0.021$ & $0.198 \pm 0.009$ & $0.219$ & $0.006$ & $14107$ & $0.16$ & $0.41$ \\
NGC1566 & 1 & $10.79$ & $-10.13$ & $0.939 \pm 0.026$ & $0.402 \pm 0.022$ & $0.562$ & $-0.126$ & $87693$ & $0.22$ & $0.69$ \\
NGC1672 & 1 & $10.73$ & $-9.85$ & $1.070 \pm 0.040$ & $0.419 \pm 0.027$ & $0.586$ & $-0.209$ & $23678$ & $0.20$ & $0.67$ \\
NGC2835 & 1 & $10.00$ & $-9.90$ & $0.601 \pm 0.093$ & $0.333 \pm 0.066$ & $0.351$ & $0.121$ & $36589$ & $0.24$ & $0.43$ \\
NGC3351 & 1 & $10.37$ & $-10.25$ & $0.553 \pm 0.076$ & $0.325 \pm 0.038$ & $0.276$ & $0.172$ & $29151$ & $0.18$ & $0.29$ \\
NGC3627 & 1 & $10.84$ & $-10.25$ & $1.064 \pm 0.024$ & $0.780 \pm 0.020$ & $0.949$ & $-0.229$ & $67195$ & $0.19$ & $0.71$ \\
NGC4254 & 1 & $10.42$ & $-9.93$ & $1.006 \pm 0.039$ & $0.601 \pm 0.026$ & $0.741$ & $-0.144$ & $65123$ & $0.15$ & $0.78$ \\
NGC4303 & 1 & $10.51$ & $-9.78$ & $0.952 \pm 0.024$ & $0.666 \pm 0.016$ & $0.879$ & $-0.170$ & $29408$ & $0.17$ & $0.69$ \\
NGC4321 & 1 & $10.75$ & $-10.20$ & $1.021 \pm 0.024$ & $0.517 \pm 0.015$ & $0.518$ & $-0.012$ & $37677$ & $0.16$ & $0.64$ \\
NGC4535 & 1 & $10.54$ & $-10.20$ & $1.129 \pm 0.058$ & $0.350 \pm 0.044$ & $0.345$ & $-0.040$ & $31296$ & $0.17$ & $0.61$ \\
NGC4826 & 2 & $10.24$ & $-10.93$ & $1.198 \pm 0.069$ & $1.082 \pm 0.038$ & $1.254$ & $-0.421$ & $22847$ & $0.12$ & $0.81$ \\
NGC5068 & 1 & $9.41$ & $-9.97$ & $0.647 \pm 0.020$ & $0.313 \pm 0.015$ & $0.382$ & $0.066$ & $67347$ & $0.24$ & $0.40$ \\
NGC7496 & 1 & $10.00$ & $-9.65$ & $1.152 \pm 0.010$ & $0.069 \pm 0.006$ & $0.347$ & $-0.331$ & $15593$ & $0.14$ & $0.70$ \\
\enddata
\tablecomments{
Columns: Cycle --- Indicates whether the galaxy was observed as part of Cycle 1 (GO 2107) or Cycle 2 (GO 3707); $\log_{10} M_\star$ --- global stellar mass [M$_\odot$]; $\log_{10} \mathrm{SFR}/M_\star$ global specific star formation rate [yr$^{-1}$]; $m$, $b$, $x_0$ --- best fit power law scaling parameters following Equation~\ref{eq:line} relating CO(2-1) to intensity in band $X$; $C_\mathrm{band}$ --- the normalization of the best-fit relation at $I_\nu^X=1$~MJy~sr$^{-1}$ [K~km~s$^{-1}$ (MJy sr$^{-1}$)$^{-1}$];
$N_{\rm pix}$ --- number of sight lines entering the analysis, where approximately four sight lines correspond to one independent measurements; $\sigma_\mathrm{pix}$ --- rms scatter in $\log_{10}$CO(2-1) intensity about the fit for all sight lines included in bins; $r$ --- rank correlation between $\log_{10}$CO(2-1) and intensity in band $X$ for all sight lines.
}
\end{deluxetable*}

\section{How to estimate CO (2-1) intensity from PAH emission maps}\label{appendix:howto}

We suggest the following recipe, along with some key caveats, to estimate CO~(2-1) intensity in units of K~km~s$^{-1}$ from JWST observations of PAH-dominated filters in units of MJy~sr$^{-1}$.

\begin{enumerate}
\item In the case of F335M or F770W, estimate and subtract the associate stellar continuum to calculate $I_{\rm F335M}^{\rm PAH}$ or $I_{\rm F770W}^{\rm PAH}$. Specifically, subtract $0.22\times I_{\rm F300M}$ or $0.13\times I_{\rm F200W}$ from $I_{\rm F770W}$ \citep[see \S\ref{subsec:jwst} and][]{sutter2024}. If this is not possible, we view F335M as not useful, while F770W can be used with our provided equations but will be biased high in regions of high stellar-to-dust ratios, including stellar bars and bulges. We do not consider stellar continuum correction necessary for F1130W.

\item For a disk galaxy with inclination $i$, correct the surface brightness by multiplying by $\cos i$.

\item If the specific star formation rate, SFR/$M_\star$, is known \citep[e.g. from ][]{leroy2019} and the galaxy resembles a low redshift star-forming galaxy (so that our corrections could be expected to apply), evaluate Equation~\ref{eq:norms} to obtain $C_{\rm norm}^{\rm F770W}$. We prefer the SFR/$M_\star$ based correction. For F335M$_\mathrm{PAH}$ and F1130W, we recommend multiplying \norm\ (top right panel of Figure~\ref{fig:co_mir_bygal}) by typical band ratios F335M$_\mathrm{PAH}$/F770W$_\mathrm{PAH} \approx 0.04$ or F1130W/F770W$_\mathrm{PAH} \approx 0.69$ (using $x_0$ from Table~\ref{tab:fits}) to yield $C_\mathrm{F335M}^\mathrm{PAH}$ and $C_\mathrm{F1130W}$ respectively. If this step is not possible, one can make a less accurate estimate by applying the general relation (i.e., simply use the appropriate equation from Table~\ref{tab:fits} and ignore the following step). In this case, one should expect $\approx 0.2$~dex bias in the predicted CO~(2-1) map for any individual galaxy.

\item If $\log_{10}C_{\rm norm}$ has been estimated, subtract this from the observed $\log_{10} \ipah$ for each sight line to remove the galaxy-to-galaxy normalization. Then for F770W$_{\rm PAH}$, plug this normalized intensity, $x$, into Equation~\ref{eq:copah_renorm} to estimate $y\equiv \log_{10}I_{\rm CO(2-1)}$. For F335M$_\mathrm{PAH}$,
\begin{equation}\label{eq:copah_renorm_3p3}
y=(0.93\pm 0.09)(x-0.10)+(0.13\pm 0.06), 
\end{equation}
where $x$ and $y$ are defined as in Equation~\ref{eq:copah_renorm} and the full Cycle 1 data set shows $0.42$~dex rms scatter in the residuals. Similarly for F1130W, we find
\begin{equation}\label{eq:copah_renorm_11p3}
y=(1.01\pm 0.09)(x-1.34)+(1.27\pm 0.08).
\end{equation}
\end{enumerate}
{The best-fit parameters to Eq.~\ref{eq:copah_renorm} for all PAH bands are shown in Table~\ref{tab:big_fit_table_normed}.
}

\begin{deluxetable*}{lrrrrrrr}
\tablecolumns{8}
\tablecaption{Best-fit parameters for $\log_{10}C$ versus global galaxy properties. The top row of Figure~\ref{fig:co_mir_all} shows the fits for F770W$_\mathrm{PAH}$.}
\label{tab:norm_fits}
\tablehead{
    \colhead{Band} & \colhead{$N_\mathrm{gal}$} & \colhead{$m$} & \colhead{$b$} & \colhead{$x_0$} & \colhead{$r$} & \colhead{$\sigma_\mathrm{line}$} & \colhead{$\sigma_\mathrm{data}$}\\
}
\startdata
\cutinhead{Fits vs $\log_{10}$ SFR/M$_\star$ [yr$^{-1}$]}
F335M$_\mathrm{PAH}$ & 19 & $-0.32 \pm 0.18$ & $1.20 \pm 0.06$ & $-10.05$ & $-0.38$ & $0.22$ & $0.20$ \\
F770W$_\mathrm{PAH}$ & 70 & $-0.21 \pm 0.04$ & $0.03 \pm 0.02$ & $-10.14$ & $-0.50$ & $0.13$ & $0.17$ \\
F770W & 70 & $-0.17 \pm 0.04$ & $-0.00 \pm 0.02$ & $-10.14$ & $-0.42$ & $0.12$ & $0.17$ \\
F1130W & 20 & $-0.01 \pm 0.13$ & $-0.03 \pm 0.05$ & $-10.08$ & $-0.01$ & $0.25$ & $0.25$ \\
\cutinhead{Fits vs $\log_{10}$ M$_\star$ [M$_\odot$]}
F335M$_\mathrm{PAH}$ & 19 & $0.44 \pm 0.10$ & $1.26 \pm 0.04$ & $10.51$ & $0.74$ & $0.14$ & $0.20$ \\
F770W$_\mathrm{PAH}$ & 70 & $0.15 \pm 0.04$ & $0.06 \pm 0.02$ & $10.34$ & $0.38$ & $0.16$ & $0.17$ \\
F770W & 70 & $0.12 \pm 0.04$ & $0.02 \pm 0.02$ & $10.34$ & $0.32$ & $0.17$ & $0.17$ \\
F1130W & 20 & $0.01 \pm 0.11$ & $-0.03 \pm 0.05$ & $10.47$ & $0.02$ & $0.25$ & $0.25$ \\
\enddata
\tablecomments{Fit results for CO(2-1) normalization vs galaxy properties. $m$ is the slope, $b$ is the intercept, $x_0$ is the pivot point, $r$ is the Pearson correlation coefficient, $\sigma_\mathrm{line}$ is the scatter about the fit, and $\sigma_\mathrm{data}$ is the scatter in the $y$ direction.}
\end{deluxetable*}

\begin{deluxetable*}{lrrrrrrrr}
\label{tab:big_fit_table_normed}
\tablecolumns{9}
\tablecaption{Ratios, correlation, and scaling relations between PAH and CO~(2-1) emission. Each section of the table reports results for a different data selection. Before fitting, the CO intensities for each galaxy were scaled by the predicted normalization based on the fit of normalization versus $\log_{10}$SFR/$M_\star$ (Fig.~\ref{fig:co_mir_bygal}, Table~\ref{tab:norm_fits}).}

\tablehead{
\colhead{$X$} & \colhead{$N_\mathrm{gal}$} & \colhead{$N_\mathrm{pix}$\tablenotemark{a}} & \colhead{$\log_{10}\mathrm{CO}/X$} & \colhead{$r$} & \colhead{$b$} & \colhead{$m$} & \colhead{$x_0$ } & \colhead{$\sigma$}}
\startdata
\cutinhead{All pixels}
F335M$_\mathrm{PAH}$ & 19 & 296834 & $0.08 \pm 0.36$ & $0.57$ & $0.13 \pm 0.06$ & $0.93 \pm 0.09$ & $0.10$ & $0.42$ \\
F770W$_\mathrm{PAH}$ & 70 & 2090731 & $-0.02 \pm 0.32$ & $0.68$ & $1.36 \pm 0.06$ & $0.88 \pm 0.06$ & $1.44$ & $0.43$ \\
F770W & 70 & 2120025 & $-0.02 \pm 0.33$ & $0.68$ & $1.41 \pm 0.07$ & $0.90 \pm 0.06$ & $1.47$ & $0.43$ \\
F1130W & 20 & 972892 & $-0.12 \pm 0.38$ & $0.67$ & $1.27 \pm 0.08$ & $1.01 \pm 0.09$ & $1.34$ & $0.46$ \\
\cutinhead{All pixels outside of centers}
F335M$_\mathrm{PAH}$ & 19 & 279910 & $0.06 \pm 0.36$ & $0.55$ & $-0.04 \pm 0.07$ & $1.00 \pm 0.11$ & $-0.05$ & $0.42$ \\
F770W$_\mathrm{PAH}$ & 70 & 2041245 & $-0.02 \pm 0.32$ & $0.67$ & $1.12 \pm 0.07$ & $0.98 \pm 0.08$ & $1.16$ & $0.41$ \\
F770W & 70 & 2069574 & $-0.02 \pm 0.33$ & $0.67$ & $1.16 \pm 0.08$ & $1.02 \pm 0.09$ & $1.19$ & $0.41$ \\
F1130W & 20 & 946437 & $-0.12 \pm 0.38$ & $0.66$ & $1.03 \pm 0.09$ & $1.14 \pm 0.14$ & $1.11$ & $0.45$ \\
\cutinhead{All pixels in centers}
F335M$_\mathrm{PAH}$ & 17 & 16915 & $0.32 \pm 0.39$ & $0.71$ & $0.35 \pm 0.07$ & $0.71 \pm 0.09$ & $0.10$ & $0.42$ \\
F770W$_\mathrm{PAH}$ & 58 & 49469 & $0.00 \pm 0.33$ & $0.88$ & $1.38 \pm 0.07$ & $0.87 \pm 0.07$ & $1.44$ & $0.41$ \\
F770W & 58 & 50434 & $-0.02 \pm 0.33$ & $0.88$ & $1.40 \pm 0.07$ & $0.91 \pm 0.07$ & $1.47$ & $0.43$ \\
F1130W & 18 & 26443 & $-0.21 \pm 0.37$ & $0.91$ & $1.15 \pm 0.07$ & $1.15 \pm 0.09$ & $1.34$ & $0.40$ \\
\cutinhead{All pixels in nebular regions (Cycle 1 only)}
F335M$_\mathrm{PAH}$ & 19 & 119051 & $0.02 \pm 0.32$ & $0.66$ & $-0.05 \pm 0.07$ & $1.03 \pm 0.12$ & $-0.05$ & $0.38$ \\
F770W$_\mathrm{PAH}$ & 19 & 186605 & $-0.12 \pm 0.32$ & $0.79$ & $0.95 \pm 0.09$ & $1.07 \pm 0.10$ & $1.09$ & $0.39$ \\
F770W & 19 & 187137 & $-0.11 \pm 0.31$ & $0.79$ & $1.00 \pm 0.08$ & $1.11 \pm 0.10$ & $1.12$ & $0.39$ \\
F1130W & 19 & 194681 & $-0.18 \pm 0.35$ & $0.79$ & $0.93 \pm 0.09$ & $1.20 \pm 0.14$ & $1.11$ & $0.41$ \\
\cutinhead{All pixels in diffuse regions (Cycle 1 only)}
F335M$_\mathrm{PAH}$ & 19 & 160845 & $0.10 \pm 0.38$ & $0.44$ & $-0.26 \pm 0.10$ & $1.24 \pm 0.30$ & $-0.40$ & $0.44$ \\
F770W$_\mathrm{PAH}$ & 19 & 632119 & $-0.02 \pm 0.35$ & $0.58$ & $0.66 \pm 0.12$ & $1.30 \pm 0.28$ & $0.66$ & $0.45$ \\
F770W & 19 & 642906 & $-0.02 \pm 0.35$ & $0.57$ & $0.70 \pm 0.12$ & $1.35 \pm 0.26$ & $0.70$ & $0.45$ \\
F1130W & 19 & 727342 & $-0.11 \pm 0.38$ & $0.58$ & $0.75 \pm 0.12$ & $1.40 \pm 0.23$ & $0.81$ & $0.46$ \\
\enddata
\tablecomments{Fit results for CO(2-1) normalization vs galaxy properties. $m$ is the slope, $b$ is the intercept, $x_0$ is the pivot point, $r$ is the Pearson correlation coefficient, and $\sigma$ is the scatter about the fit.}

\end{deluxetable*}

This procedure will predict the CO~(2-1) intensity. In many cases the molecular gas mass will be the quantity of direct interest. In future work, we aim to explore how \ipah\ compares to $N(H)$ directly. For now, to convert from CO intensity to molecular gas column or surface density, one should multiply the predicted CO maps by an appropriate CO~(2-1)-to-H$_2$ conversion factor. \citet{bolatto2013} review the CO-to-H$_2$ conversion factor and \citet{schinnerer2024} provide current prescriptions that attempt to account for variations in opacity, CO excitation, and metallicity.

\bibliography{wise_co_paper_v2}{} %
\bibliographystyle{aasjournal}

\end{document}